\documentclass[a4paper]{article}

\usepackage[dvips]{graphicx}

\usepackage[a4paper]{geometry}
\usepackage{algorithm}
\usepackage{algpseudocode}
\usepackage{amsmath}
\usepackage{mathtools}
\usepackage{cite}      
\usepackage{color}

\newcommand{\dif}{{\rm d}}

\newcommand{\dvol}{{\rm d}^3{\bf r}}
\newcommand{\dsur}{{\rm d}^2{\bf r}}

\begin{document}

\title{\flushleft \bf Electromagnetic modelling of superconductors with a smooth current-voltage relation:\\
variational principle and coils from a few turns to large magnets\footnote{{\color{blue} Final version published as E Pardo, J \v Souc and L Frolek 2015 Supercond. Sci. Technol. {\bf 28} 044003, doi:10.1088/0953-2048/28/4/044003. Several minor typos have been corrected in the published version.}}}

\author{Enric Pardo, J\'an \v Souc, Lubomir Frolek\\
{\normalsize Institute of Electrical Engineering, Slovak Academy of Sciences,}\\
{\normalsize Dubravska 9, 84104 Bratislava, Slovakia}\\
{\normalsize enric.pardo@savba.sk}
}

\date{\today}

\maketitle

\begin{abstract}
Many large-scale applications require electromagnetic modelling with extensive numerical computations, such as magnets or 3-dimensional (3D) objects like transposed conductors or motors and generators. Therefore, it is necessary to develop computationally time-efficient but still accurate numerical methods. This article develops a general variational formalism for any ${\bf E}({\bf J})$ relation and applies it to model coated-conductor coils containing up to thousands of turns, taking magnetization currents fully into account. The variational principle, valid for any 3D situation, restricts the computations to the sample volume, reducing the computation time. However, no additional magnetic materials interacting with the superconductor are taken directly into account. Regarding the coil modelling, we use a power law $E(J)$ relation with magnetic field-dependent critical current density, $J_c$, and power law exponent, $n$. We test the numerical model by comparing the results to analytical formulas for thin strips and experiments for stacks of pancake coils, finding a very good agreement. Afterwards, we model a magnet-size coil of 4000 turns (stack of 20 pancake coils of 200 turns each). We found that the AC loss is mainly due to magnetization currents. We also found that for an $n$ exponent of 20, the magnetization currents are greatly suppressed after 1 hour relaxation. In addition, in coated conductor coils magnetization currents have an important impact on the generated magnetic field; which should be taken into account for magnet design. In conclusion, the presented numerical method fulfills the requirements for electromagnetic design of coated conductor windings.
\end{abstract}




\section{Introduction}

Recently, there have been important advances in superconducting large-scale or power applications, partly thanks to the development of ReBCO coated conductors\footnote{ReBCO stands for $Re$Ba$_2$Cu$_3$O$_{7-x}$, where $Re$ is a rare earth, typically Y, Gd or Sm.} and the maturity of MgB$_2$ wires. An important issue in these applications is the electromagnetic design, implying quantities such as the AC loss and, for magnets, the magnetic field quality. Several problems require extensive numerical computations \cite{reviewac}, such as coated conductor magnets containing thousands of turns \cite{neighbour} or 3-dimensional (3D) problems such as wires or cables \cite{nii12SST,amemiya14SST,grilli13Cry,zermeno13SST,stenvall14SST} or motors and generators \cite{masson13IES,zermeno14SST}, among others. Therefore, fast and efficient but still accurate numerical methods for complex situations are required. This work is intended to develop a general variational formalism, including 3-dimensional situations, and apply it to numerically calculate a complex system like magnets with thousands of turns.

In the past, there have been significant efforts to compute 3D situations by directly solving a master partial differential equation using finite element methods (FEM) \cite{reviewac}, although for all cases the spatial discretization is relatively coarse, due to the required long computing time. Most of the FEM methods (and the totality of those used in commercial software) need to set the boundary conditions far away from the sample, requiring a high number of elements in the air and increasing the computing time. In order to overcome this problem, the coupled boundary-element/finite-element method (BEM-FEM) was developed \cite{russenschuck99rep,kurz02IES} and showed good results to model generated magnetic fields with superconducting magnets with a ferromagnetic yoke \cite{kurz00IES,hagen12IES}. However, this method has not been applied to electromagnetic time-evolution problems, such as relaxation effects in the superconductor and AC loss. In addition, integral formulations of the ${\bf T}$ current potential have also been shown to reduce the computation region to the superconductor \cite{nii12SST}, although this feature is only applicable for thin films (including surfaces with 3D bending in complicated structures like Roebel cables).

Calculations based on variational principles have been shown to be time-efficient, due to both optimized numerical routines and reducing the computation volume to the superconducting region(s) \cite{roebelcomp}. Although it has been shown that for mathematically 2D problems (infinitely long or cylindrical shapes or flat thin films) the region of study can be restricted to the sample volume \cite{prigozhin96JCP,prigozhin97IES,prigozhin98JCP,sanchez01PRB,navau01PRB,HacIacinphase,pancaketheo}, there has not been published any variational principle with this property in 3D. Nevertheless, there have been pioneering contributions to 3D variational principles in the ${\bf H}$ formulation, which requires setting boundary conditions far away from the sample; and thence, requiring elements in the air \cite{bossavit94IEM,badia02PRB,badia12SST}. For all formulations developed up to present, variational principles have limitations to describe ferromagnetic materials interacting with the superconductor. These require to be linear, with either arbitrary \cite{pancakefm} or infinite \cite{sanchez10APLb} permeability. For simplicity, this article regards only the situations with no ferromagnetic materials.

An additional feature of the general 3D problem is to determine a realistic ${\bf E}({\bf J})$ relation of the superconductor (where ${\bf E}$ is the electrical field and ${\bf J}$ is the current density) for ${\bf J}$ with a component in the magnetic field direction (flux cutting situation), which causes non-parallel ${\bf E}$ and ${\bf J}$. Although there have been interesting theoretical \cite{clem11PRB,badia09PRB} and experimental \cite{clem11SST} works on this issue, the ${\bf E}({\bf J})$ relation with non-parallel ${\bf E}$ and ${\bf J}$ remains mostly unknown. The present work does not investigate this problem, allowing any ${\bf E}({\bf J})$ relation as input.

Independently to the development of numerical models, the computation of the AC loss in coated conductor coils have been an active field of study, by either using variational principles \cite{pancaketheo,pancakefm,pardo12SSTb,neighbour,prigozhin11SST} or solving differential equations by FEM \cite{grilli07SST,ainslie11SST,zermeno11IES,zhangM12APL,zhangM14SST,guC13IES,gomory13IES,vetrella14IES}. However, most of the works only regard single pancake coils (or pancakes) or stacks of few pancakes. In \cite{pardo12SSTb,zhangM14SST}, stacks of many pancakes have been studied but with few turns in each pancake. Nevertheless, \cite{neighbour} calculated the AC loss in a magnet-size coil of 4000 turns (200 turns per pancake), although for the sharp ${\bf E}({\bf J})$ relation of the critical state (figure \ref{f.EJCSM}) with constant $J_c$. However, magnet design (and other power applications) require magnetic field-dependent $J_c$ and smooth ${\bf E}({\bf J})$ relation, the latter being essential to investigate relaxation effects.

In this article, we obtain a variational principle for 3D bodies that restrict the calculation volume to the sample (section \ref{s.func}). In that section, we also regard infinitely long or cylindrical symmetries. Afterwards (section \ref{s.nummeth}), we detail the numerical method for circular coils to minimize the functional from the variational principle obtained in the previous section. In section \ref{s.expmeth}, we present the measurement technique used to obtain the input data (the critical current density $J_c$ as a function of the magnetic field and its orientation, and the power-law exponent $n$ of the power-law $E(J)$ relation) and the AC loss, which is compared to the measurements. The next step is to benchmark the numerical method by comparing the calculations to the strip formulas and the experiments on coils made of a few pancakes (section \ref{s.valid}). Once the model is tested, we investigate a magnet-size coil (20 pancakes of 200 turns, totaling 4000 turns) for a smooth $E(J)$ relation and a magnetic field-dependent $J_c$ and discuss the main features of the electromagnetic response, regarding relaxation after the application of a DC input current, cyclic input current, and the effect of magnetization currents on the generated magnetic field (section \ref{s.magnet}). Finally, we present our conclusion (section \ref{s.conc}).


\section{Variational principle for 3D bodies}
\label{s.func}

In the following, we present the 3D variational principle for any material with non-linear ${\bf E}({\bf J})$ relation, such as superconductors. The nathematical method is based on calculating the current density ${\bf J}$ and scalar potential $\phi$ (or magnetic field ${\bf H}$) by minimizing a certain functional. First, we obtain the functional in the ${\bf H}$ formulation and afterwards in the ${\bf J}-\phi$ one. In order to give a certain name, we call the variational principle and the numerical method to solve it presented in this article as Minimum Electro-Magnetic Entropy Production (MEMEP), since the solution minimizes the entropy production due to the electromagnetic fields, as discussed in section \ref{s.thermo}.

Actually, there have been several important contributions to the field. First, Bossavit found the 3D functional for the ${\bf H}$ formulation for any ${\bf E}({\bf J})$ relation, including the multi-valued ${\bf E}({\bf J})$ relation of the critical-state model \cite{bossavit94IEM}. However, some key steps in the deduction are omitted in his article. Later on, Badia and Lopez provided a physical insight of the functional and applied the Euler-Lagrange formalism \cite{badia12SST}, although this mathematical framework is strictly only valid for smooth ${\bf E}({\bf J})$ relations. In addition, the ${\bf H}$ formulation has the handicap that the boundary conditions for general sample shapes need to be set far away from the sample, requiring unnecessary elements in the air. Prigozhin introduced the ${\bf J}$ formulation, where the volume of study is reduced to the sample volume, although only for mathematically two-dimensional shapes \cite{prigozhin96JCP,prigozhin97IES,prigozhin98JCP}.

This section not only presents a comprehensive deduction of the ${\bf H}$ formulation directly from Maxwell equations with a material ${\bf E}({\bf J})$ relation but also introduces the ${\bf J}-\phi$ formulation for the general 3D case.


\subsection{${\bf H}$ formulation}

\begin{figure}[tbp]
\begin{center}
\includegraphics[width=8cm]{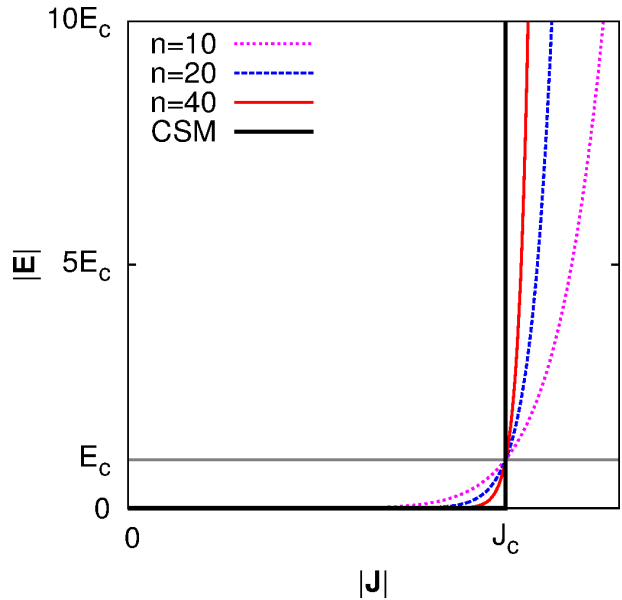}%
\caption{\label{f.EJCSM} Qualitative features of the ${\bf E}({\bf J})$ relations for the isotropic power-law ${\bf E}=E_c(|{\bf J|}/J_c)^n({\bf J}/|{\bf J}|)$ and critical-state model (CSM). Bossavit's ${\bf E}({\bf J})$ relation in \cite{bossavit94IEM} actually assumes that the material is linear for $|{\bf E}|$ above a certain threshold.}
\end{center}
\end{figure}

Our goal is to find the functional that by minimizing it, we obtain the solution for ${\bf H}$ (in general, the solution of ${\bf H}$ corresponds to an extreme: minimum, maximum or saddle). Its Euler-Lagrange equations (see \ref{s.EL} on the Euler-Lagrange equations of a functional) should be Faraday's law for a certain ${\bf E}({\bf J})$ relation
\begin{equation}
\label{faraday}
\mu_0{\dot {\bf H}}+\nabla\times{\bf E}({\bf J})=0,
\end{equation}
where we assume that there are no magnetic materials, ${\bf B}=\mu_0{\bf H}$, and ${\dot {\bf H}}\equiv\partial{\bf H}/\partial t$, with $t$ being the time. Using $\nabla\times{\bf H}={\bf J}$, which corresponds to Ampere's law with negligible displacement current\footnote{The influence of the displacement current, $\partial {\bf D}/\partial t$, on the current distribution in closed current loops (or multi-turn coils) is negligible for conductor lengths, $l$, much shorter than the radiation wavelength $\lambda$ \cite{johnson}. This can be regarded as a rule of thumb for any situation, since in magneto-statics the current always forms a closed loop. Setting a stricter criterion than for antenna design, $l<\lambda/100$ instead of $l<\lambda/10$ (section 5-1 of \cite{johnson}), the displacement current does not influence the current density for frequencies up to around 3 MHz and 3 kHz for conductor lengths of 1 m and 1 km, respectively. Even when the current density in the conductor is not influenced by the displacement current, there will still be a certain small radiation power loss due to the oscillating magnetic dipole moment. This loss is $P_r=\sqrt{\mu_0/\epsilon_0}[4\pi^3N^2A^2I_m^2/(3\lambda^4)]$; where $N$ is the number of turns in the coil, $A$ is the area of one turn, and $I_m$ is the current amplitude (equation (5-5) of \cite{johnson}). However, this contribution is typically negligible compared to the non-linear Joule AC loss in superconductors.}, we obtain the differential equation for ${\bf H}$,
\begin{equation}
\label{faradayH}
\mu_0{\dot{\bf H}}+\nabla\times{\bf E}(\nabla\times{\bf H})=0.
\end{equation}
We also assume a layered discretization of time. That is, we approximate the time derivatives as ${\dot {\bf H}}\approx\Delta{\bf H}/\Delta t$ with the same time interval for all positions and $\Delta{\bf H}$ is the variation of ${\bf H}$ between two time layers, for instance $t_0$ and $t_0+\Delta t$. Then, Faraday's law becomes
\begin{equation}
\label{faradayDH}
\mu_0\frac{\Delta{\bf H}}{\Delta t}+\nabla\times{\bf E}(\nabla\times{\bf H}_0+\nabla\times\Delta{\bf H})=0,
\end{equation}
where ${\bf H}_0$ is the magnetic field at the beginning of the interval, $t_0$, while the time at the end is $t_0+\Delta t$. Similarly, the functional and the Euler-Lagrange equations take the form
\begin{equation}
\label{LHgen}
L=\int_V{\rm d}^3{\bf r}f({\bf r},\Delta H_i({\bf r}),\partial_j\Delta H_i({\bf r}))
\end{equation}
and
\begin{equation}
\label{ELH}
\frac{\partial f}{\partial \Delta H_i}-\sum_{j=1}^3\partial_j\left [ \frac{\partial f}{\partial(\partial_j\Delta H_i)} \right ]=0,
\end{equation}
respectively, where $V$ is any 3D volume, ${\bf r}=(x_1,x_2,x_3)\equiv(x,y,z)$ and $i\in\{x,y,z\}$. Below, we omit the upper limit of the sums, with the understanding that it is 3. Equations (\ref{faradayH}) and (\ref{ELH}) are conveniently separated into two terms, one depending only on $\Delta H_i$ and another depending on its spacial partial derivatives. Using that $\nabla\times\Delta{\bf H}=\sum_{kji}\epsilon_{kji}\partial_j\Delta H_i{\bf e}_k$, where $\epsilon_{kji}$ is the antisymmetric Levi-Civita symbol and ${\bf e}_k$ is the unit vector in direction $k$, equation (\ref{ELH}) becomes
\begin{equation}
\label{ELHJ}
\frac{\partial f}{\partial H_i}+\sum_{jk}\epsilon_{ijk}\partial_j\left [ \frac{\partial f}{\partial J_k} \right ]=0.
\end{equation}
Then, if the functional is of the form
\begin{eqnarray}
L & = & \int_V{\rm d}^3{\bf r}f(\Delta H_i({\bf r}),\partial_j\Delta H_i({\bf r}))  \nonumber \\
& = & \int_V\dvol \left[{ \frac{1}{2}\mu_0\frac{(\Delta {\bf H})^2}{\Delta t}+U({\bf J}_0+\Delta {\bf J}) }\right], \label{fHgen}
\end{eqnarray}
where ${\bf J}_0=\nabla\times{\bf H}_0$ is the current density at the beginning of the time layer, $\Delta{\bf J}=\nabla\times\Delta {\bf H}$ is the variation of ${\bf J}$ at the end of the time layer, and $U({\bf J})$ is a function such that ${\bf E}=\nabla_{\bf J}U$, being $\nabla_{\bf J}$ the gradient with respect to ${\bf J}$, the Euler-Lagrange equation (\ref{ELHJ}) becomes Faraday's equation, as expressed in (\ref{faradayDH}). Next, we show that for any physical ${\bf E}({\bf J})$ relation there exists a single scalar function defined as
\begin{equation}
\label{UJ}
U({\bf J})=\int_0^{\bf J}{\bf E}({\bf J'})\cdot{\rm d}{\bf J'}.
\end{equation} 
First, $\nabla_{\bf J}\times{\bf E}$ follows
\begin{equation}
\nabla_{\bf J}\times{\bf E}({\bf J})=\sum_{ijk}\epsilon_{ijk}\frac{\partial E_k({\bf J})}{\partial J_j}{\bf e}_i=\sum_{ijk}\epsilon_{ijk}\rho_{kj}({\bf J}){\bf e}_i.
\end{equation}
Second, from thermodynamical principles, it can be shown that the differential resistivity matrix $\tilde\rho$, with components $\rho_{kj}$, is symmetric and positive definite \cite{badia12SST}. As a consequence, $\sum_{jk}\epsilon_{ijk}\rho_{kj}=0$, and thence
\begin{equation}
\nabla_{\bf J}\times{\bf E}=0.
\end{equation}
Therefore, from Stokes' theorem it follows that $\oint{\bf E}({\bf J}')\cdot{\rm d}{\bf J}'=0$; and hence the scalar function $U({\bf J})$ from equation (\ref{UJ}) is well defined because it does not depend on the integration path. In addition Bossavit found that the $U({\bf J})$ function above can also be applied to the multi-valued ${\bf E}({\bf J})$ relation of the critical state model \cite{bossavit94IEM} (see sketch of the ${\bf E}({\bf J})$ relation for isotropic cases, ${\bf E}\parallel{\bf J}$, in figure \ref{f.EJCSM}).

Next, we regard the case that the superconductor is submitted to a certain given applied magnetic field ${\bf H}_a$. In that case there are two contributions to the magnetic field, the applied magnetic field and the magnetic field created by the current density, ${\bf H}_J$; and thence ${\bf H}={\bf H}_J+{\bf H}_a$. The functional (\ref{LHgen}) with the functional density (\ref{fHgen}) becomes
\begin{equation}
\label{LHa}
L=\int_V\dvol\left [ \frac{1}{2}\mu_0\frac{(\Delta{\bf H}_J)^2}{\Delta t} + \mu_0\Delta{\bf H}_a\frac{\Delta{\bf H}_J}{\Delta t} + U({\bf J}_0+\Delta{\bf J}) \right ],
\end{equation}
where we dropped the term $(1/2)\mu_0(\Delta{\bf H}_a)^2$ because it does not depend on the minimization variable $\Delta{\bf H}_J$.

One possibility to impose a current constraint, that is the total transport current $I$ is given, is to construct an augmented functional with a Lagrange multiplier $\lambda$, 
\begin{equation}
L_a=L+\lambda\left [ \int_{S_1}({\bf J}_0+\Delta{\bf J})\cdot{\rm d}{\bf s}-I \right ]^2=L+\lambda\left [ \oint_{\partial S_1}({\bf H}_0+\Delta{\bf H})\cdot{\rm d}{\bf l}-I \right ]^2,
\end{equation}
where $L$ is the functional in (\ref{LHa}), $S_1$ is any cross-section where the net current is $I$, $\partial S_1$ is its boundary, ${\rm d}{\bf s}$ is the surface differential and ${\rm d}{\bf l}$ is the path differential. Since we are using that ${\bf J}=\nabla\times{\bf H}$, this implies $\nabla\cdot{\bf J}=0$, and thence there is current continuity in the whole body. For multiply connected superconductors, such as multi-tape cables or coils, one can add as many additional terms in the functional as current constraints. In the minimization of the augmented functional above, the Lagrange multiplier $\lambda$ should be treated as an independent variable.


\subsection{${\bf J}-\phi$ formulation}
\label{s.Jphi}

Next, we deduce an equivalent 3D functional for a formalism depending on the current density ${\bf J}$ and the scalar potential $\phi$. This ${\bf J}-\phi$ formulation allows to greatly reduce the number of variables compared to the $\bf H$ formulation, since the computation volume is restricted to the superconductor. Actually, the ${\bf J}$ formulation also requires computing the scalar potential $\phi$ (or the electric charge density $q$) but the important reduction of variables in the air justifies adding this additional scalar field. In the ${\bf J}$ formulation, we may regard either ${\bf J}$ or the vector potential ${\bf A}$ as state variables. 

First, we wish to find a fundamental equation for ${\bf A}$ and $\phi$. For this purpose, we take the relation between ${\bf E}$ and the potentials
\begin{equation}
\label{EAphi}
{\bf E}+\dot{\bf A}+\nabla\phi=0.
\end{equation}
Given a certain ${\bf E}({\bf J})$ relation and using that for Coulomb's gauge, $\nabla\cdot{\bf A}=0$, the Ampere's law $\nabla\times{\bf H}={\bf J}$ becomes\footnote{Again, we neglect the displacement current ($\partial{\bf D}/\partial t\approx 0$) and assume no magnetic materials (${\bf B}=\mu_0{\bf H}$). We also use the definition of vector potential ${\bf B}=\nabla\times{\bf A}$.} $\nabla^2{\bf A}=-\mu_0{\bf J}$. Note that in Coulomb's gauge the scalar potential becomes the electrostatic potential (see Appendix B of \cite{reviewac}). Then, we obtain the system of equations
\begin{eqnarray}
{\bf E}(-\nabla^2{\bf A}/\mu_0)+\dot{\bf A}+\nabla\phi=0, \label{EAphiCou} \\
\nabla\cdot{\bf A}=0, \label{Cougauge}
\end{eqnarray}
where the Coulomb's gauge condition is added to the initial relation (\ref{EAphi}). The reason is that (\ref{EAphi}) consists of 3 equations (one for each vector component), while there are 4 fields to be determined (the 3 components of ${\bf A}$ and $\phi$). In addition, the Poisson equation for ${\bf A}$, $\nabla^2{\bf A}=-\mu_0{\bf J}$, does not directly imply the continuity equation for ${\bf J}$, $\nabla\cdot{\bf J}=0$. Explicitly, $\mu_0\nabla\cdot{\bf J}=-\nabla\cdot(\nabla^2{\bf A})=\nabla\cdot[\nabla\times(\nabla\times{\bf A})-\nabla(\nabla\cdot{\bf A})]=-\nabla^2(\nabla\cdot{\bf A})$, and thence $\nabla\cdot{\bf J}=0$ follows for $\nabla\cdot{\bf A}=0$ but not for any gauge.

Next, we consider a layered time discretization, so that $\dot{\bf A}({\bf r})\approx\Delta{\bf A}({\bf r})/\Delta t$ with the same $\Delta t$ for any $\bf r$ and where $\Delta {\bf A}({\bf r})$ is the variation of ${\bf A}({\bf r})$ between time $t=t_0$ and $t=t_0+\Delta t$. Then, equations (\ref{EAphiCou}) and (\ref{Cougauge}) become
\begin{eqnarray}
{\bf E}[-\nabla^2({\bf A}_0+\Delta {\bf A})/\mu_0]+\frac{\Delta{\bf A}}{\Delta t}+\nabla\phi=0, \label{EDAphi} \\
\nabla\cdot({\bf A}_0+\Delta {\bf A})=0, \label{CouDA}
\end{eqnarray}
where ${\bf A}_0\equiv{\bf A}(t=t_0)$ and we used the fact that equations (\ref{EAphiCou}) and (\ref{Cougauge}) hold for all the previous time layers, and therefore $\nabla\cdot{\bf A}_0=0$.

In the following, we present a certain functional and then we proof that its Euler-Lagrange equations are (\ref{EDAphi}) and (\ref{CouDA}). The ansatz of the functional with functional density $f$ is
\begin{eqnarray}
L & = & \int_V\dvol f \label{funcAphi} \\
& = & \int_V\dvol \left[{ \frac{1}{2}\frac{\Delta{\bf A}}{\Delta t}\cdot\Delta{\bf J} + U({\bf J}_0+\Delta{\bf J})+\nabla\phi\cdot({\bf J}_0+\Delta{\bf J}) }\right] \label{fDA}
\end{eqnarray}
with ${\bf J}_0=-\nabla^2{\bf A}_0/\mu_0$, $\Delta{\bf J}=-\nabla^2\Delta{\bf A}/\mu_0$ and $U({\bf J})$ defined as equation (\ref{UJ}), which follows ${\bf E}=\nabla_{\bf J}U$. Since the functional above depends on the second space derivative of $\Delta{\bf A}$ through $\Delta{\bf J}$, its Euler-Lagrange equations also contain second derivatives (see section \ref{s.EL}) as
\begin{eqnarray}
\label{ELDA}
\frac{\partial f}{\partial \Delta A_i}-\sum_j\partial_j\left [ \frac{\partial f}{\partial(\partial_j\Delta A_i)} \right]+
\sum_{jk} \partial_j\partial_k \left [ \frac{\partial f}{\partial(\partial_j\partial_k\Delta A_i)} \right ]=0 \\
\frac{\partial f}{\partial \phi}-\sum_j\partial_j\left [ \frac{\partial f}{\partial(\partial_j\phi)} \right]=0.
\end{eqnarray}
The equation for $\phi$ results in 
\begin{equation}
\label{DJ0}
\nabla\cdot({\bf J}_0+\Delta{\bf J})=0.
\end{equation}
Using $\nabla^2{\bf A}=-\mu_0{\bf J}$ and the vector relation $\nabla^2{\bf A}=\nabla\times(\nabla\times{\bf A})-\nabla(\nabla\cdot{\bf A})$, equation (\ref{DJ0}) turns into
\begin{equation}
\label{D2DA}
\nabla^2[\nabla\cdot({\bf A}_0+\Delta{\bf A})]=0.
\end{equation}
Appliying the Euler-Lagrange equations (\ref{ELDA}) to the functional density we obtain
\begin{equation}
\label{D2EAphi}
\nabla^2\left [{ \frac{\Delta{\bf A}}{\Delta t}+{\bf E}({\bf J}_0+\Delta{\bf J})+\nabla\phi }\right ]=0.
\end{equation}
In order to deduce the equation above we used that, since $\nabla^2\Delta{\bf A}=-\mu_0\Delta{\bf J}$, then 
\begin{equation}
\frac{\partial f}{\partial(\partial_j\partial_k\Delta A_i)}=-\frac{\partial f}{\partial J_i}\frac{\delta_{jk}}{\mu_0},
\end{equation}
where $\delta_{jk}$ is 1 when $j=k$ and 0 otherwise. Therefore, we have obtained that the Euler-Lagrange equations from the functional density (\ref{fDA}), equations (\ref{D2DA}) and (\ref{D2EAphi}), correspond to (\ref{CouDA}) and (\ref{EDAphi}) with a global Laplacian operator. Actually, for a general 3D body the part within the Laplacian of (\ref{D2DA}) and (\ref{D2EAphi}) also vanishes, obtaining equations (\ref{CouDA}) and (\ref{EDAphi}). This is because for any scalar or vector function, for instance $g({\bf r})$ and ${\bf G}({\bf r})$, respectively, the fact that its Laplacian is zero implies $g({\bf r})=0$ and ${\bf G}({\bf r})=0$, as long as those functions are also zero at the boundaries of the volume where their Laplacian vanishes. For finite 3D bodies, the potentials ${\bf A}$ and $\nabla{\phi}$ approach zero at infinity. Since we are neglecting electromagnetic radiation, ${\bf E}$ also vanishes far away from the sample. For the idealization of infinitely long wires or cables transporting a certain net current, the wire or cable actually contains a returning conductor that closes the circuit (see section \ref{s.long} for details). As a consequence, all fields actually vanish at infinity and equations (\ref{D2DA}) and (\ref{D2EAphi}) imply (\ref{CouDA}) and (\ref{EDAphi}), respectively.

Then, we have found that the $\Delta{\bf A}$ and $\phi$ that correspond to an extreme of the functional (\ref{funcAphi}) are the solutions of the magnetostatic problem. For the cases that $\nabla \phi$ is given by an external source, such as circular coils or long conductors, it can be proofed that the extreme is a minimum (see \ref{s.convex}). For the general case, it is not clear to the authors whether the extreme is always a minimum. Notice that for a mathematical method that finds the extreme of this functional by changing $\Delta{\bf A}$ and $\phi$, one should take those fields into account for the whole 3D space, or for a volume much larger than the sample volume where it is possible to set known boundary conditions. Although that is feasible by allowing the void to be conductive with a large resistivity, and thence allowing a residual ${\bf J}$ outside the sample; we are departing from the goal of reducing the computational volume. This can be solved as follows.

Actually, we can also find the extreme of the functional by using $\Delta{\bf J}$ and $\phi$ instead of $\Delta{\bf A}$ and $\phi$, as long as we keep the condition $\nabla^2\Delta{\bf A}=-\mu_0\Delta{\bf J}$. The solution of $\Delta{\bf A}$ for this Poisson equation is \cite{Jackson}
\begin{equation}
\label{DAJ}
\Delta{\bf A}({\bf r})=\frac{\mu_0}{4\pi}\int_V\dvol ' \frac{\Delta{\bf J}({\bf r}')}{|{\bf r}-{\bf r}'|}.
\end{equation}
Notice that from the equation above, $\nabla\cdot\Delta{\bf A}=0$ only when ${\nabla}\cdot\Delta{\bf J}=0$. Then, by using the integral equation above, the functional density of (\ref{fDA}) only depends on $\Delta{\bf J}$. Moreover, if the functional is at an extreme for a certain $\Delta{\bf J}$, it will also be at a extreme with the corresponding $\Delta{\bf A}$ from equation (\ref{DAJ}), and thence the electromagnetic quantities follow equations (\ref{EDAphi}) and (\ref{CouDA}), as well as $\nabla\cdot{\bf J}=0$. Now, the boundary conditions for $\Delta{\bf J}$ can be directly set on the sample surface, and thence the integration volume in the functional ({\ref{funcAphi}) can be restricted to the sample volume. However, the $\Delta{\bf A}$ that $\Delta{\bf J}$ generates extends to the whole 3D space.

In case that the sample is submitted to a given applied magnetic field ${\bf H}_a=\nabla\times{\bf A}_a/\mu_0$, where ${\bf A}_a$ is the applied vector potential generated by external currents of given magnitude\footnote{The magnetic field generated by any magnetic material may also be reagarded as that generated by an equivalent magnetizaton current density $\nabla\times{\bf M}$, where ${\bf M}$ is the magnetization.} ${\bf J}_a$, the vector potential may be separated into two contributions, one from the current density in the sample ${\bf J}$ (or $\Delta{\bf J}$) and one from the applied field conribution, and thence ${\bf A}={\bf A}_J+{\bf A}_a$ and $\Delta{\bf A}=\Delta{\bf A}_J+\Delta{\bf A}_a$. For this case, one should use an expression for ${\bf A}_a$ that follows $\nabla\cdot{\bf A}_a=0$. Then, the functional becomes
\begin{equation}
\label{LDAa}
L=\int_V\dvol \left [ { \frac{1}{2}\frac{\Delta{\bf A}_J}{\Delta t}\cdot\Delta{\bf J}+\frac{\Delta{\bf A}_a}{\Delta t}\cdot\Delta{\bf J}+U({\bf J}_0+\Delta{\bf J})+\nabla\phi\cdot({\bf J}_0+\Delta{\bf J}) } \right ],
\end{equation}
where $\Delta{\bf A}_J$ is related to $\Delta{\bf J}$ by equation (\ref{DAJ}) and we have used that $\int_V\dvol\Delta{\bf A}_J\cdot\Delta{\bf J}_a=\int_V\dvol\Delta{\bf J}\cdot\Delta{\bf A}_a$. In the functional above, we have also dropped the terms with $\Delta{\bf A}_a\cdot\Delta{\bf J}_a$ and $\nabla\phi\cdot{\bf J}_a$, since these quantities are fixed (note that in the term with $\nabla\phi\cdot{\bf J}_a$, $\phi$ refers to the scalar potential in the region where ${\bf J}_a$ is flowing).

The boundary conditions for $\Delta{\bf J}$ are the following. For finite 3D samples under an applied magnetic field, one may simply impose that the current does not flow outwards from the sample, and thence $\Delta{\bf J}\cdot{\bf e}_n=0$, where ${\bf e}_n$ is the unit vector perpendicular to the surface. For transposed infinitely long wires and cables, it is necessary to take a periodicity condition into account, in order to reduce the problem to one transposition length or a fraction of it. In the case that there is a certain given transport current $I$, an additonal constraint on $\Delta{\bf J}$ should be imposed, as follows.

One option is to set the current constraints is by an augmented functional, such as
\begin{equation}
L_a=L+\lambda\left [ \int_{S_1}({\bf J}_0+\Delta{\bf J})\cdot{\rm d}{\bf s}-I \right ]^2,
\end{equation}
where $L$ is the functional of (\ref{LDAa}), $S_1$ is any cross-section that transports the current $I$, ${\rm d}{\bf s}$ is the surface differential, and $\lambda$ is a Lagrange multiplier that has to be treated as an independent variable. However, our numerical method for minimization for coils (section \ref{s.genmeth}) takes implicitly the current constraint, and thence an augmented functional is not necessary.

In principle, for general 3D problems one may have difficulties setting the boundary conditions for $\phi$ (or $\nabla\phi$) on the sample surface. If this problem arises, it may be solved by using that, since we are using Coulomb's gauge, $\phi$ is the electrostatic potential, and thence it is related to the surface and volume charge densities, $\sigma$ and $q$ respectively, as
\begin{equation}
\phi({\bf r})=\frac{1}{4\pi\epsilon_0} \left [ { \int_V\dvol ' \frac{q({\bf r}')}{|{\bf r}-{\bf r}'|} + \oint_{\partial V} {\rm d}{s}\frac{\sigma({\bf r}')}{|{\bf r}-{\bf r}'|} } \right ],
\end{equation}
where $\partial V$ is the volume surface and ${\rm d}{s}$ is its differential. With this approach, the variables are $\Delta{\bf J}$, $q$ and $\sigma$, which are all constricted within the sample volume.


\subsection{Thermodynamical interpretation}
\label{s.thermo}

Previously, Badia and Lopez provided a thermodynamical interpretation for situations close to the critical-state model using the ${\bf H}$ formulation \cite{badia02PRB,badia12SST}. In the following, the extend and detail the analysis for any ${\bf E}({\bf J})$ relation, also for the ${\bf J}-\phi$ formulation. 

First, we outline a description on the energetic meaning of the several terms of the functionals (\ref{fHgen}) and (\ref{fDA}) in the ${\bf H}$ and ${\bf J}-\phi$ formulations, respectively. For this purpose, we take into account that for the limit of $\Delta t\to 0$, we obtain $U({\bf J}_0+\Delta{\bf J})-U({\bf J}_0)\approx{\bf E}_0\cdot\Delta{\bf J}$, where ${\bf E}={\bf E}(t_0)$. Then, the functionals become, save a constant term with $U({\bf J}_0)$,
\begin{equation}
\label{LDHDt}
L\approx\frac{1}{\Delta t}\int_V\dvol \left[{ \frac{1}{2}\mu_0(\Delta{\bf H})^2+\Delta t\ {\bf E}_0\cdot\Delta{\bf J} }\right]
\end{equation}
for the ${\bf H}$ formalism and
\begin{equation}
L\approx\frac{1}{\Delta t}\int_V\dvol \left[{ \frac{1}{2}\Delta{\bf A}\cdot\Delta{\bf J}+\Delta t\ {\bf E}_0\cdot\Delta{\bf J}+\Delta t\ \nabla\phi\cdot({\bf J}_0+\Delta{\bf J}) }\right] \label{LDJDt}
\end{equation}
for the ${\bf J}-\phi$ one. The first term in (\ref{LDHDt}) is the magnetic energy of the magnetic field variation $\Delta{\bf H}$ ignoring the interaction with the pre-existing magnetic field ${\bf H}_0$, while the second term is twice the heat generated during the time interval $\Delta t$ due to the onset of $\Delta{\bf J}=\nabla\times\Delta{\bf H}$ (here we use that for the first Taylor approximation $\Delta{\bf J}$ increases linearly with time). Regarding the functional in the ${\bf J}-\phi$ formulation, the second term is identical to the ${\bf H}$ formulation and the first term is, similarly, the magnetic energy of $\Delta J$ ignoring the presence of the pre-existing current density ${\bf J}_0$. The third term is twice the energy transferred to the electrostatic system, as long as $\bf A$ is in Coulomb's gauge; and thence $\phi$ is the electrostatic potential. Although there are strong similarities with the Lagrangian formalism of classical mechanics, as Badia and Lopez showed in \cite{badia12SST}, this analogy is incomplete because the first term is not the total energy of the system. Therefore, it is not clear that the functionals in our system can be interpreted as Lagrangians in the classical sense.

In the following, we investigate the resulting functional from minimizing the magnetic variable, $\Delta{\bf H}$ or $\Delta{\bf J}$, and its interpretation. Regarding the ${\bf J}-\phi$ formulation, the system follows the Euler-Lagrange equations of the functional, which correspond to (\ref{EAphi}) and (\ref{DJ0}). By using these equations and the relation $\nabla\phi\cdot{\bf J}=\nabla\cdot(\phi{\bf J})-\phi\nabla\cdot{\bf J}$, the minimized functional becomes
\begin{eqnarray}
L_{\rm min} & = & \int_V \dvol \left[{ \frac{1}{2} {\bf E}_0\cdot\Delta{\bf J} + \frac{1}{2} \nabla\cdot(\phi{\bf J}) }\right]  \nonumber \\
& = & \frac{1}{2} \int_V \dvol {\bf E}_0\cdot\Delta{\bf J} + \frac{1}{2} \oint_{\partial V} \dif{\bf s}\cdot{\bf J}\phi, \label{Lminfull}
\end{eqnarray}
where in the last step we used Stokes Theorem. In equation (\ref{Lminfull}) above, the first term is the average heat rate generation due to the onset of $\Delta{\bf J}$ during the time interval $\Delta t$ and the second one is the energy rate flowing outwards the sample. If we include all sources of current in the system, as we have done in the previous sections, the second term drops. Then,
\begin{equation}
\label{Lmin}
L_{\rm min} \approx \int_V \dvol \frac{1}{2} {\bf E}_0\cdot\Delta{\bf J}.
\end{equation}
Following a similar argument, the functional for the ${\bf H}$ formalism in (\ref{LDHDt}) results also in (\ref{Lmin}). Therefore, the electromagnetic solution obtained from minimizing both functionals is identical. Notice that $L_{\rm min}$ from the equation above is the average rate of heat generation between time $t_0$ and $t_0+\Delta t$ due to $\Delta{\bf J}$. This is because for a superconductor ${\bf E}\cdot{\bf J}$ is the density of local heat rate generation, as justified in \cite{reviewac}. The total rate of heat generation is
\begin{eqnarray}
{\dot Q} & = & \int_V \dvol \left[{ \frac{1}{2} {\bf E}_0\cdot\Delta{\bf J} + {\bf E}_0\cdot{\bf J}_0 }\right] \nonumber \\
& = & L_{\rm min}+\int_V \dvol \ {\bf E}_0\cdot{\bf J}_0, \label{Qprod}
\end{eqnarray}
resulting in a heat variation
\begin{equation}
\delta Q=L_{\rm min}\delta t+\delta t\int_V \dvol \ {\bf E}_0\cdot{\bf J}_0.
\end{equation}
Since from the second law of thermodynamics $\delta Q=T\delta S$, where $T$ is the temperature and $\delta S$ is the entropy, then the rate of entropy production is
\begin{equation}
{\dot S}=\frac{1}{T}\left[{ L_{\rm min}+\int_V \dvol \ {\bf E}_0\cdot{\bf J}_0 }\right].
\end{equation}
Since the second term does not depend on $\Delta{\bf J}$, we have found that for isothermal conditions (as it is usually assumed for purely magnetic modelling) the $\Delta{\bf J}$ that minimizes the functional $L$ also minimizes the rate of entropy production. In addition, the rate of heat production, equation (\ref{Qprod}), is also minimum. This suggests that the functionals from (\ref{LDHDt}) and (\ref{LDJDt}) may correspond to the entropy production, save constant terms.


\subsection{Long straight wires and cables}
\label{s.long}

In this section, we present the modifications to the functional in the ${\bf J}-\phi$ formulation for infinitely long straight wires or cables (referred below as ``conductors") transporting a certain current $I$.

Let us take $z$ as the direction that the conductor extends infinitely. Then, the current density and vector potential follow the $z$ direction and do not depend on $z$; and thence $\Delta{\bf J}({\bf r})=\Delta J(x,y){\bf e}_z$ and $\Delta{\bf J}({\bf r})=\Delta J(x,y){\bf e}_z$, where $x$ and $y$ are the other Cartesian components and ${\bf e}_z$ is the unit vector in the $z$ direction. For this case, the magnetic induction ${\bf B}$ is perpendicular to ${\bf J}$, and thence there is no flux cutting. As a result, ${\bf E}$ is parallel to ${\bf J}$ and ${\bf E}({\bf J})=E(J){\bf e}_z$, where $E$ has the same sign as $J$. Then from ${\bf E}({\bf r})=E(x,y){\bf e}_z$ and equation (\ref{EAphi}) follows that $\nabla\phi({\bf r})=\partial_z\phi{\bf e}_z$, which is constant within the conductor (or each tape or filament in multi-tape or multi-filament conductors). In addition, the function for $U$ in (\ref{UJ}) can be simplified as $U(J)=\int_0^J E(J')\dif J'$. Then, the functional in (\ref{LDAa}) becomes
\begin{eqnarray}
L & = & l\int_S\dsur f \nonumber \\
& = & l\int_S\dsur \left[{ \frac{1}{2}\frac{\Delta A_J}{\Delta t}\Delta J+\frac{\Delta A_a}{\Delta t}\Delta J+U(J_0+\Delta J)+\partial_z\phi({\bf J}_0+\Delta{\bf J}) }\right], \label{Linf}
\end{eqnarray}
where $l$ is the conductor length, $S$ is the superconductor cross-section in the $xy$ plane, and $\dsur$ is $\dif x\dif y$. 

Since any conductor transporting a certain current should form a closed circuit, we may consider a returning conductor  separated by a certain distance $D$ much larger than both the conductor width and thickness but still much shorter than the conductor length $l$. The functional of this system is 
\begin{equation}
\label{Linfret}
L=l\int_{S_+}\dsur f+l\int_{S_-}\dsur f,
\end{equation}
where $f$ is the functional density of equation (\ref{Linf}), $S_+$ is the section of the conductor transporting current $I$ (or ``main" conductor) and $S_-$ is the returning conductor, with transport current $-I$. 

Next, we pay attention to the ``main" conductor only. The variation of vector potential has two components, $\Delta A=\Delta A_J+\Delta A_{\rm int}$, regarding the contribution from $\Delta J$ in the ``main" conductor and the interaction with the returning one. By direct integration of (\ref{DAJ}) for a wire of length $l$ without adding any additional constant\footnote{Far away from a wire of arbitrary cross-section and arbitrary internal distribution of current, the vector potential is the same as an infinitesimally thin wire.}, $A_{\rm int}\approx-\mu_0I\ln(l/D)/(2\pi)$. It is important to notice that the interaction term $A_{\rm int}$ is constant. Therefore, $\Delta J$ in the ``main" conductor is independent on $\Delta J$ in the returning one, as long as the total current is fixed. As a consequence, the two terms of the functional (\ref{Linfret}) can be minimized independently. For the ``main" wire, the functional turns into
\begin{eqnarray}
L & = & l\int_{S_+}\dsur \left[{ \frac{1}{2}\Delta A_J\Delta J+\Delta A_a\Delta J+U(J_0+\Delta J)+\partial_z\phi({\bf J}_0+\Delta{\bf J}) }\right] \nonumber \\
& & -l\frac{\mu_0}{4\pi}(\Delta I)^2\ln\left( \frac{l}{D} \right).
\end{eqnarray}
For minimization purposes, the last term is constant and could be dropped, as well as the general $l$ factor.


\subsection{Axi-symmetric systems and coils}
\label{s.cylsym}

This section obtains a simplified functional for the ${\bf J}-\phi$ formalism valid for axi-symmetric systems, also regarding multiple connected bodies made of concentric rings.

For bodies with axial symmetry, ${\bf J}$ (and $\Delta{\bf J}$) follow the angular direction and do not depend on the angular coordinate $\varphi$. Therefore, $\Delta{\bf J}({\bf r})=\Delta J(r,z){\bf e}_\varphi$ and $\Delta{\bf A}({\bf r})=\Delta A(r,z){\bf e}_\varphi$, where $r$ and $z$ are the radial and axial components and ${\bf e}_\varphi$ is the unit vector in the angular direction. As for infinitely long conductors, axi-symmetric superconductors do not present flux cutting and ${\bf E}({\bf J})=E(J){\bf e}_\varphi$, where $E$ has the same sign as $J$. Then, the function for $U$ in (\ref{UJ}) becomes simply $U(J)=\int_0^J E(J')\dif J'$. Finally, as a consequence of (\ref{EAphi}) and the axial symmetry, $\nabla\phi(r,z)=(1/r)\partial_\varphi\phi{\bf e}_\varphi$ and $\partial_\varphi\phi$ is constant within each isolated ring (or each turn separately for a coil with axial symmetry).

Circular coils may be approximated as a set of concentric rings with given current $I$. The drop of electrostatic potential that drives the current in each turn may be regarded either as a separate voltage source in each turn or, more realistically, as a global source but with turns that break the circular symmetry in only one point in order to connect with its neighbour turn (figure \ref{f.sketchcoil}).

\begin{figure}[tbp]
\begin{center}
\includegraphics[width=6cm]{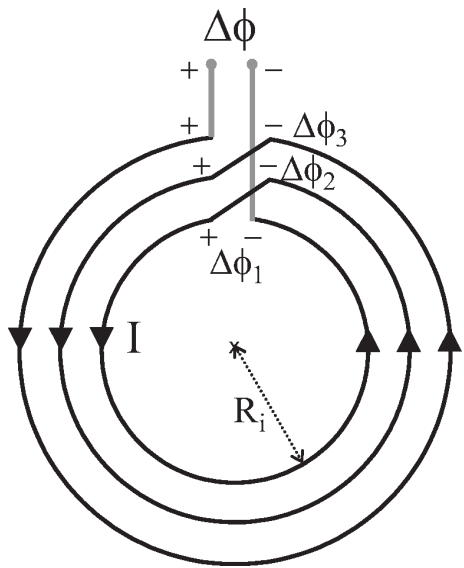}%
\caption{\label{f.sketchcoil} Closely packed pancake coils can be well approximated as an axi-symmetric problem. The turns are assumed circular, except at a small section where it connects with the following turn. At that point the voltage drop at a certain turn $i$ is defined $\Delta \phi_i$.}
\end{center}
\end{figure}

Therefore, the functional of (\ref{LDAa}) becomes
\begin{eqnarray}
\label{Lcyl}
L=2\pi\sum_{i=1}^{n_t} \bigg( & \int_{S_i} & {\rm d}{s} \ r\left [{ \frac{1}{2}\frac{\Delta A_J}{\Delta t}\Delta J+\frac{\Delta A_a}{\Delta t}\Delta J+U(J_0+\Delta J) }\right ]  \nonumber \\
& + & \partial_\varphi\phi_i\int_{S_i}{\rm d}s({\bf J}_0+\Delta{\bf J}) \bigg),
\end{eqnarray}
where $n_t$ is the number of simply connected regions (or number of turns in coils), ${\rm d}s$ is ${\rm d}r{\rm d}z$, $S_i$ is the surface cross-section of region $i$, and $\partial_\varphi\phi_i$ is $\partial_\varphi\phi$ at the same region $i$. In principle, the current constraint may be set by Lagrange multipliers, as described in section (\ref{s.Jphi}). However, our minimization process maintains a constant net current, only considering variations of $\Delta J$ that do not modify the net current. Therefore, the last term in (\ref{Lcyl}) becomes $\partial_\varphi\phi_i I_i$, where $I_i$ is the net current. This term does not depend on the particular distribution of $\Delta J$, and thence it may be dropped from the functional, resulting in
\begin{equation}
\label{Lcyl}
L=2\pi \int_{S} {\rm d}{s} \ r\left [{ \frac{1}{2}\frac{\Delta A_J}{\Delta t}\Delta J+\frac{\Delta A_a}{\Delta t}\Delta J+U(J_0+\Delta J) }\right ],
\end{equation}
where all surface integrals are merged into one in order to simplify the notation.


\subsection{Magnetic field-dependent or position-dependent ${\bf E}({\bf J})$ relation}

The electromagnetic problem can be solved by minimizing the functional in (\ref{LDAa}) also for magnetic field-dependent and position-dependent ${\bf E}({\bf J})$ relations.

In practice, the ${\bf E}({\bf J})$ relation depends on the magnetic field\footnote{Actually, ${\bf B}$ is the magnetic flux density. However, since we assume no magnetic materials, ${\bf B}=\mu_0{\bf H}$, and thence $\bf B$ and $\bf H$ play the same physical role, except of a constant. Therefore, in this article we name both as ``magnetic field" for simplicity.} $\bf B$; such as an isotropic power-law relation ${\bf E}=E_c(|{\bf J}|/J_c)^n({\bf J}/|{\bf J}|)$ with magnetic field-dependent parameters $J_c({\bf B})$ and $n({\bf B})$, and constant $E_c$. This case can be solved iteratively as follows. First, ${\bf B}$ is taken as that at $t=t_0$. Then, $\Delta{\bf J}$ and $\phi$ are solved by minimizing (\ref{LDAa}). Afterwards, ${\bf B}$ is calculated again and the process is repeated until the difference in $\Delta {\bf J}$ and $\phi$ between two iterations is below a certain tolerance (more details in section \ref{s.EJBmeth}).

For a position-dependent ${\bf E}({\bf J})$ relation, ${\bf E}({\bf J},{\bf r})$, one simply has to take the corresponding position-dependent $U({\bf J},{\bf r})=\int_0^{\bf J}\dif{\bf J}'\cdot{\bf E}({\bf J}',{\bf r})$. An example of these ${\bf E}({\bf J},{\bf r})$ relations are isotropic power laws with position-dependent $J_c$ or $n$, caused by are either non-uniform material properties \cite{jiangZ07PhC,solovyov13SST} or thickness variations in thin films \cite{tsukamoto05SST}.


\section{Numerical method for coils}
\label{s.nummeth}

In this section, we present the details of the numerical method to obtain the time evolution of $\bf J$ in a superconducting coil, and from this result calculate the rest of the electromagnetic parameters, such as the generated magnetic field, the critical current, the AC loss, and the coil voltage.

In short, given the current density for a certain time $t_0$, ${\bf J}_0$, the method finds the current density ${\bf J}={\bf J}_0+\Delta{\bf J}$ for a time $t=t_0+\Delta t$. This is done by minimizing the functional (\ref{LDAa}) by keeping the total transport current in the coil constant. We also assume an axi-symmetric symmetry of the coil, as outlined in section \ref{s.cylsym}. However, the method presented here can also be applied to long conductors by taking a closed ring loop of the conductor and setting an average radius much larger than the conductor width and thickness. 


\subsection{$E(J)$ relation}

Although the numerical method presented here is valid for any $E(J)$ relation, the results in this work are for a power-law expression as \cite{brandt95RPP,plummer87IEM,rhyner93PhC}
\begin{equation}
\label{EJ}
E(J)=E_c\left( \frac{|J|}{J_c} \right)^n \frac{J}{|J|},
\end{equation}
where the parameters are the critical current density $J_c$, the power-law exponent $n$ and the voltage criterion for the critical current density $E_c$. The relation above is generally valid for superconductors with $J$ close to the critical current density \cite{brandt95RPP}. With this ${\bf E}({\bf J})$ relation, the function $U({\bf J})$ in equation (\ref{UJ}) becomes
\begin{equation}
U(J)= \frac{1}{n+1} E_cJ_c\left( \frac{|J|}{J_c} \right)^{n+1},
\end{equation}
where $J$ is the (only) axial component of ${\bf J}$.

In general, the parameters $J_c$ and $n$ of the power law depend of the magnetic field $\bf B$. In this article, we use $J_c({\bf B})$ and $n({\bf B})$ dependencies extracted from measurements (see sections \ref{s.Jcn} and \ref{s.JcBmagnet}). In a similar way, the numerical method is prepared to take position-dependent parameters, $J_c(r,z)$ and $n(r,z)$, into account; although we do not present results in this work.


\subsection{Discretization}

In order minimize the functional and find the solution of the current density, we divide the entire cross-section into $N$ elements where we assume uniform current density. The computations in this article are done with a uniform mesh discretization with elements of rectangular cross-section. However, the method is also valid for elements of any cross-section shape, such as triangular ones, at least when the cross-sectional area of all elements is the same \cite{ruuskanen14prp}. The formalism below is written taking this into account. The reason to use a uniform mesh is to minimze the RAM memory (see the last paragraph of this section for details).

Given a certain element $i$, we define $I_i(t)$ as the current in that element at a time $t$. Then, the current density at that element is $J_i=I_i/S_i$, where $S_i$ is the cross-section of element $i$. In consistency with the notation in section (\ref{s.Jphi}), we denote $I_{0,i}=J_{0,i}/S_i$ and $\Delta I_i=\Delta J_i/S_i$, where $I_{0,i}$ and $\Delta I_i$ are the current in element $i$ at the previous time, $t_0$, where $J$ is solved and $\Delta I_i$ is the change in current in element $i$ when increasing the time by $\Delta t$. We also define the average magnetic flux in element $i$ cross-section as 
\begin{equation}
F_i=\frac{2\pi}{S_i}\int_{S_i}{\rm d}s\ rA(r,z).
\end{equation}
Similarly, quantities $\Delta F_{J,i}$ and $\Delta F_{a,i}$ are defined as $F_i$ but replacing $A$ by $\Delta A_J$ and $\Delta A_a$, respectively. With our discretization, $\Delta F_i$ from the definition above and equation (\ref{DAJ}) becomes
\begin{equation}
\Delta F_{J,i}=\sum_iC_{ij}\Delta I_i,
\end{equation}
where the sum is done for all elements, $1 \le i \le N$, and the constant terms $C_{ij}$ are
\begin{equation}
\label{cij}
C_{ij}=\frac{2\pi}{S_iS_j}\int_{S_i}\dif s \int_{S_j}\dif s' ra_{\rm loop}(r,r',z-z').
\end{equation}
In the equation above, ${\bf r}$ and ${\bf r}'$ are 3D vector positions, while $r$ and $r'$ are the radial components only, and $a_{\rm loop}$ is the vector potential generated by a circular loop per unit current in the loop with radius $r'$ located at height $z'$. The expression of this function is given by equations (\ref{aloop}) and (\ref{k}) in \ref{s.intmat}. The matrix elements $C_{ij}$ are numerically evaluated, as detailed in \cite{pancaketheo}. Using the $C_{ij}$ matrix, the functional in (\ref{Lcyl}) for our discretization becomes
\begin{equation}
L=\frac{1}{2\Delta t}\sum_{ij}C_{ij}\Delta I_i\Delta I_j+\frac{1}{\Delta t}\sum_i \Delta F_{a,i}\Delta I_i + 2\pi\sum_ir_iS_iU_i,
\end{equation}
where $U_i\equiv U[(I_{0,i}+\Delta I_i)/S_i]$. Next, for minimization purposes, we regard a change in $L$ due to a change $\delta I$ of $\Delta I_i$. The resulting change in $L$ is
\begin{eqnarray}
\delta L_i & = & \frac{1}{\Delta t}(\Delta F_{J,i}+\Delta F_{a,i})\delta I+\frac{1}{2\Delta t}C_{ii}(\delta I)^2 \nonumber \\
& & +2\pi r_iS_i \left [{ U\left( \frac{I_{0,i}+\Delta I_i+\delta I}{S_i} \right) - U_i }\right ].
\end{eqnarray}
For the rest of this article, the quantities between braces refer to the vector composed by the value of that quantity among all elements, such as $\{I_i\}$, $\{\Delta F_i\}$ or $\{U_i\}$.

Since we take a magnetic field-dependent $J_c$ and $n$ into account, we also need to compute the magnetic field in the superconductor. This is done numerically, as follows. For our discretization, we take the average magnetic field in a given element $i$ cross-section as the relevant quantity to calculate $J_c({\bf B}$) and $n({\bf B})$ in that element, that is 
\begin{equation}
{\bf B}_i\equiv\frac{1}{S_i}\int_{S_i}\dif s {\bf B}({\bf r}).
\end{equation}
Accordingly, we define ${\bf B}_{J,i}$ and ${\bf B}_{a,i}$ by substituting ${\bf B}$ in the equation above by the magnetic field created by the currents and the applied field, ${\bf B}_J$ and ${\bf B}_a$, respectively. Since ${\bf B}_a$ is usually analytical, it can be straightforwardly calculated. By means of the Biot-Savart law, ${\bf B}_{J,i}$ can be calculated as the sum of the contributions from all elements as
\begin{equation}
{\bf B}_{J,i}=\sum_i{\bf b}_{ij}I_j
\end{equation}
with
\begin{equation}
\label{bij}
{\bf b}_{ij}=\frac{1}{S_iS_j}\int_{S_i} \dif s \int_{S_j} \dif s' {\bf b}_{\rm loop}(r,r',z-z'),
\end{equation}
where ${\bf b}_{\rm loop}$ is the magnetic field created by an infinitely thin circular loop per unit current in the loop, given by equations (\ref{brloop})-(\ref{k}) in \ref{s.intmat}. That appendix also presents the numerical method to evaluate ${\bf b}_{ij}$.

For meshes with rectangular elements of identical cross-section, the independent entries of the interaction matrices $C_{ij}$, $b_{r,ij}$ and $b_{z,ij}$ can be greatly reduced comparing to arbitrary non-uniform meshes. For the latter, the each interaction matrix contains $N^2$ independent entries, where $N$ is the total number of elements. For the former, the number of independent entries in the matrices can be reduced as follows. From equations (\ref{cij}) and (\ref{bij}) we can see that, $C_{ij}$, $b_{r,ij}$ and $b_{z,ij}$ only depend on $r_i$, $r_j$ and $z_i-z_j$, where $(r_i,z_i)$ and $(r_j,z_j)$ are the coordinates of the cross-sectional center of elements $i$ and $j$, respectively. For given a $r_i$ and $r_j$, the number of different $z_i-z_j$ is only $n_z(2n_p-1)$, where $n_z$ and $n_p$ are the number of elements in the $z$ direction in each single tape and are the number of pancakes, respectively. Then, the number of independent entries in each interaction matrix is $n_r^2n_z(2n_p-1)$, which is much smaller than the one for the general case $N^2=(n_rn_zn_p)^2$, where $n_r$ is the number of elements in the $r$ direction. As a result, uniform mesh allows a reduction in RAM memory storage or the interaction matrices, which occupies most of the memory storage of the program, by a factor $n_zn_p/(2-1/n_p)$. For $n_z=100$, this reduction is by a factor around 100 and 1000 for 1 and 20 pancakes, respectively.


\subsection{General minimization method}
\label{s.genmeth}

As mentioned above, our minimization method contains the current constraints as a built-in feature, and therefore minimization with Lagrange multipliers is not necessary.

The main steps of the minimization process are the following (see algorithm \ref{a.genmeth}). We start with a physical current distribution, $\{I_i\}$, corresponding to a certain time $t_0$, and assign 0 to $\{\Delta I_i\}$. Then, we set the net transport current in each turn by distributing the change in transport current, $\Delta I_{\rm tran}$, uniformly over the cross-section of each turn. Next, we find the induced magnetization currents, as follows. For each turn, we find the element $i_+$ where adding a certain positive value $\delta I=h$ to $\Delta I_{i_+}$ decreases $L$ the most (or increases it the least). In this routine, the value of $h$ sets the tolerance. We continue by finding the element $i_-$ within the same turn where substracting $h$ (or adding $\delta I=-h$) to $\Delta I_{i_-}$ minimizes the most $L$. We do the same operation for all turns, so that we find the pair $i_+$ and $i_-$ that minimizes the most $L$ for the whole coil, taking elements $i_+$ and $i_-$ that belong to the same turn. Thus, we reduce $L$ while keeping the current constraint. Once we have found the optimum pair, we set the new values to the pair, $\Delta I_{i_+}\gets \Delta {I_{i_+}}+h$ and $\Delta I_{i_-}\gets \Delta {I_{i_-}}-h$. Afterwards, we update $\{\Delta F_i\}$ and $\{U_i\}$ accordingly. In this way, we do not need to evaluate these quantities at each evaluation of $\delta L$, only when the change in current is finally set. We continue this process until any pair of elements with $\delta I=+h$ and $\delta I=-h$ will increase $L$ instead of decreasing it. This minimization routine will always obtain the current distribution corresponding to the global minimum of $L$ within the tolerance $h$, as detailed in \cite{HacIacinphase}.

\begin{algorithm}
\caption{The minimization method in pseudo-code below rapidly calculates the current distribution in the coil by minimizing the functional in (\ref{Lcyl}). More details are provided in the text.}\label{a.genmeth}
\begin{algorithmic}[l]
\State Set $\Delta I_i \gets 0$ for all $i$;
\State Add $\Delta I_{\rm tran}$ distributed uniformly among all elements;
\Repeat
	\State $\delta L\gets 1$;	
	\For{${\rm turn}=1$ to $n_t$}
		\State Find the element $i_+$ within the present turn
		\State where {\em adding} $h$ to $\Delta I_{i_+}$ produces the smallest $\delta L_{i_+}$;
		\State Find the element $i_-$ within the present turn
		\State where {\em substracting} $h$ from $\Delta I_{i_-}$ produces the smallest $\delta L_{i_-}$;	
		\If{ $\delta L>\delta L_{i_+}+\delta L_{i_-}$ }		
			\State $\delta L\gets \delta L_{i_+}+\delta L_{i_-}$;
			\State ${i_{+,{\rm min}}}\gets i_+$;
			\State ${i_{-,{\rm min}}}\gets i_-$;
		\EndIf
	\EndFor
	\If{ $\delta L < 0$ }
		\State $\Delta I_{i_{+,{\rm min}}}\gets \Delta I_{i_{+,{\rm min}}}+h$;
		\State $\Delta I_{i_{-,{\rm min}}}\gets \Delta I_{i_{-,{\rm min}}}-h$;
		\State Update $F_i$ and $U_i$ for all $i$;
	\EndIf

\Until{$\delta L\ge 0$;}
\end{algorithmic}
\end{algorithm}


\subsection{Method for magnetic field-dependent $E(J)$ relation}
\label{s.EJBmeth}

For a magnetic field-dependent $E(J)$, such as a power-law with $J_c({\bf B})$ and $n({\bf B})$, we use an iterative method as follows (see algorithm \ref{a.JcBmeth}). First, we add the change in transport current $\Delta I_{\rm tran}$ uniformly among each turn. Afterwards, we find the current distribution $\{\Delta I_i\}$ that minimizes the functional while maintaining the value of transport current, as detailed in section \ref{s.genmeth}. In order to avoid oscillations between iterations, we apply a damping factor to $\{\Delta I_i\}$, such as
\begin{equation}
\Delta I_i\gets \Delta I_{p,i}+(\Delta I_{i}-\Delta I_{p,i})K_d,
\end{equation}
where $\Delta I_{p,i}$ is the change in current of element $i$ at the previous iteration and $K_d$ is the damping factor. We found an optimum value of $K_d=0.9$ regarding computing time. Afterwards, we calculate both components of the magnetic field in all the elements, $\{B_{r,i}\}$ and $\{B_{z,i}\}$, and update the vectors containing $J_c$ and $n$, $\{J_{c,i}\}$ and $\{n_i\}$. Finally, we need to update $U$ for each element, $\{U_i\}$, as a consequence of the local change of $J_c$ and $n$. We repeat the iterations until the change of $\Delta I_i$ is below the tolerance $h$ for any element $i$.

The speed of the algorithm have been increased as follows. We start with an initial $h$ much larger than our final tolerance goal as $h=2^kh^*$, where $h^*$ is our tolerance goal and $k$ is an integer larger than one. Afterwards, we repeat the minimization process but with half the previous $h$, until $h=h^*$, our desired value. We have found an optimum exponent of $k=5$. We choose $h^*=J_{c0}S_{\rm av}/m$, where $J_{c0}=J_c(B=0)$, $S_{\rm av}$ is the average cross-section of the elements, and $m$ is an integer number ranging from 500 to 45000, being the largest values for the lowest current amplitudes. Computations for lower current amplitudes require more strict tolerance $h^*$ because, for the same time step, the average change in the current density decreases with the current amplitude, and thence the same absolute error $h^*$ corresponds to a higher relative error.

\begin{algorithm}
\caption{The interative method below in pseudo-code obtains the current distribution for an $E(J)$ relation with magnetic field-dependent parameters, such as the critical current density $J_c$ and the power-law exponent $n$.}\label{a.JcBmeth}
\begin{algorithmic}[l] 
\State Set $\Delta I_i \gets 0$ for all $i$;
\State Add $\Delta I_{\rm tran}$ distributed uniformly among all elements;
\Repeat 
	\State Set $\Delta I_{p,i}\gets \Delta I_{i}$ for all $i$;
	\State Find $\{\Delta I_i\}$ that minimizes functional
  \State while keeping the value of transport current;
	\State Set $\Delta I_i\gets \Delta I_{i,p}+(\Delta I_{i}-\Delta I_{p,i})K$ for all $i$;
	\State Calculate $\{B_{r,i}\}$ and $\{B_{z,i}\}$;
	\State Update $\{J_{c,i}\}$ and $\{n_i\}$;
	\State Update $\{U_{i}\}$;
\Until{change in $\Delta I_i$ below tolerance for any $i$;}
\State Set $I_i\gets I_i+\Delta I_i$ for all $i$;
\end{algorithmic}
\end{algorithm}


\subsection{AC loss calculation}

Once $J$ is known, the instantaneous power loss can be simply evaluated as
\begin{equation}
P=\int_V\dvol\ {\bf E}({\bf J})\cdot{\bf J}=2\pi\int_S\dif s\ rE(J)J,
\end{equation}
since ${\bf E}\cdot{\bf J}$ describes the local heat generation \cite{reviewac}. Thus, the loss per cycle (or heat generated per cycle) is simply 
\begin{equation}
Q=\oint_T\dif t\int_V\dvol\ {\bf E}\cdot{\bf J}=2\pi\oint_T\dif t\int_S\dif s\ rE(J)J,
\end{equation}
where $T$ is the period of the external excitation. For any transport current or applied magnetic field with a symmetrical waveform, such as triangular or sinusoidal, the integral can be reduced to half a period. In this article, we assume sinusoidal transport currents, $I(t)=I_m\sin(2\pi\nu t)$, of a given frequency $\nu$. For this case, we calculate the ac loss over the half cycle after the first instant that $I=-I_m$. This is because for conductors submitted to simultaneous alternating current and magnetic field, such as in coils, the loss signal becomes periodic after that instant \cite{HacIacinphase}.

Alternatively, the AC loss per cycle could be calculated from the power delivered from the power source \cite{reviewac}
\begin{equation}
Q=\oint_T\dif t\ v\cdot I,
\end{equation}
where $v$ is the voltage in the coil, calculated as detailed in section $\ref{s.vphi}$. Notice that $v\cdot I$ is not necessarily the instantaneous power dissipation in the coil, due to inductive effects.

In this article, we typically divide the AC cycle into 80 equal time steps for AC loss calculations.


\subsection{Implementation and computing times}

The numerical implementation is programmed in Fortran 95; although C++, MATLAB or other general-purpose programming languages may be used. The computations in this article have been done with either a table computer with Intel(R) Core(TM) i7-3770K CPU processor and 8 GB RAM or a server with two Intel(R) Xeon(TM) E5645 processors and 48 GB RAM, both presenting similar computing time. The program also executes computations corresponding to different current amplitudes in parallel so that all cores of the multi-core processors are simultaneously running. The computing time (using the table computer) strongly depends on the required accuracy. For the experimental stack of 4 pancake coils with 24 turns each (96 turns in total) the computing time for 50 elements in the tapes and 40 time steps per cycle is 75 min for the whole AC loss curve with 8 amplitudes, corresponding to less than 10 minutes on average per amplitude. The estimated error by comparing to the results for 100 elements per tape and 320 time steps per cylce is below 4 \%.


\subsection{Voltage and scalar potential}
\label{s.vphi}

Once the current density is known, we can evaluate the voltage drop along each turn in the whole coil as follows.

The voltage drop in any situation, including varying magnetic fields, is the difference of the electrostatic potential at the wire ends. Thus, the voltage drop at a certain turn $i$ is $v_i=-2\pi \partial_\varphi\phi_i$, where the sign of the voltage is taken in such a way that the voltage decreases when moving in the direction of the current flow and the current is defined positive if it circulates anti-clockwise. Thus, the total voltage drop in the coil is
\begin{equation}
v=\sum_{i=1}^{n_t}v_i=-2\pi\sum_{i=1}^{n_t} \partial_\varphi\phi_i.
\end{equation}

Next, we obtain $\partial_\varphi\phi_i$ from the current density from (\ref{EAphi}),
\begin{equation}
\partial_\varphi\phi=-r\left [{ E(J)+{\dot A}_J+{\dot A}_a }\right ].
\end{equation}
This defines a $\partial_\varphi\phi$ for each position in the tape cross-section. Since $\partial_\varphi\phi$ is actually constant in each turn, it is sufficient to take any arbitrary point in a turn cross-section. However, one may take an average across the turn cross-section in order to minimize the effect of any numerical errors
\begin{equation}
\partial_\varphi\phi_i=-\frac{1}{S_i}\int_{S_i}{\rm d}s\ r\left [{ E(J)+{\dot A}_J+{\dot A}_a }\right ].
\end{equation}
The applied vector potential $A_a$ is typically an analytical function, and thence its time derivative can be calculated straightforwardly. We numerically evaluate ${\dot A}_J$ at a certain time $t_0$, from $A_J$ at $t_0$ and the previous and following time layers, $t_0-\Delta t'$ and $t_0+\Delta t$ respectively, as follows
\begin{equation}
{\dot A}_J(t_0)\approx \frac{1}{2}\frac{A_J(t_0+\Delta t)-A_J(t_0)}{\Delta t}+\frac{1}{2}\frac{A_J(t_0)-A_J(t_0-\Delta t')}{\Delta t'}.
\end{equation}



\section{Experimental method}
\label{s.expmeth}

This section outlines the experimental method for the measurement of the critical current density, the flux creep exponent and the AC loss.

\subsection{Critial current density and flux-creep exponent}
\label{s.Jcn}

In this article, we use a ReBCO coated conductor tape from SuperPower \cite{SuperPower} for all experiments. This tape is 4 mm wide, with a total of 40 $\mu$m copper stabilizer layers, a 1 $\mu$m thick superconducting layer, and a self-field critical current at 77 K of 128 A.

We measured the dependence of the critical current density $J_c$ on the magnetic field magnitude $|{\bf B}|\equiv B$ and its orientation $\theta$ (see sketch in figure \ref{f.nBth}) at 77 K, as detailed in \cite{coatedIc}. In order to extract $J_c$ from measurements of the tape critical current, $I_c$, we corrected the spurious effects of the self-field, following the method in \cite{coatedIc}. The reader can find the $I_c$ measurements and extracted $J_c$ for the tape used in this article in \cite{pardo12SSTb}. For completeness, we include the extracted $J_c(B,\theta)$ relation, being
\begin{eqnarray}
J_c(B,\theta,J) = [ J_{c,ab}(B,\theta,J)^m+J_{c,c}(B)^m ]^{1/m} \label{Jcall}
\end{eqnarray}
with
\begin{eqnarray}
J_{c,ab}(B,\theta,J) = \frac{J_{0,ab}}{\left[{1+\frac{Bf(\theta,J)}{B_{0,ab}}}\right]^{\beta_{ab}}}, \label{Jcab} \\
J_{c,c}(B) = \frac{J_{0,c}}{\left[{1+\frac{B}{B_{0,c}}}\right]^{\beta_{c}}}, \label{Jc} \\
\end{eqnarray}
and
\begin{eqnarray}
f(\theta,J) = \left \{ \begin{array}{ll} 
                              f_{0}(\theta) & {\rm if}\ J\sin\theta > 0 \\
															f_{\pi}(\theta) & {\rm otherwise} 
															\end{array} \right . , \label{f} \\
f_{0}(\theta) = \sqrt{u^2\cos^2(\theta+\delta_{0})+\sin^2(\theta+\delta_{0})}, \label{f0} \\
f_{\pi}(\theta) = \sqrt{u^2\cos^2(\theta+\delta_{\pi})+v^2\sin^2(\theta+\delta_{\pi})}, \label{fpi}
\end{eqnarray}
where the parameters are $m=8$, $J_{0,ab}=2.53\cdot 10^{10}$ A/m$^2$, $J_{0,c}=2.10\cdot 10^{10}$ A/m$^2$, $B_{0,ab}=414$ mT, $B_{0,c}=90$ mT, $\beta_{ab}=0.934$, $\beta_c=0.8$, $u=5.5$, $v=$1.2, $\delta_0$=-2.5$^{\rm o}$ and $\delta_\pi$=0.5$^{\rm o}$. The critical current density at zero local field is $J_c(B=0)=2.59\cdot 10^{10}$ A/m$^2$. The estimated error of the extracted $J_c(B,\theta)$ is below 5\% \cite{pardo12SSTb}.

\begin{figure}[tbp]
\begin{center}
\includegraphics[width=9cm]{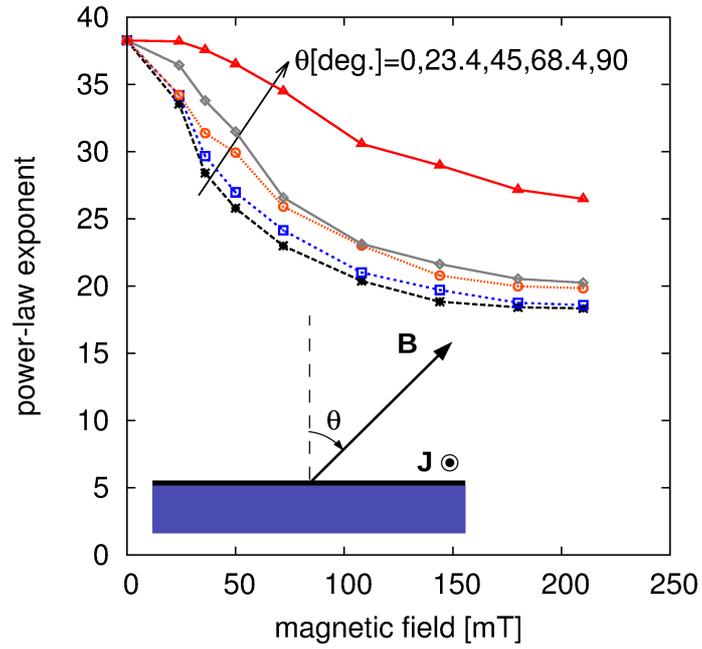}%
\caption{\label{f.nBth} Measured power-law exponent $n$ [see $E(J)$ relation in (\ref{EJ})] directly obtained from tape critical-current measurements. The insert shows a sketch of the angle $\theta$ definition, where the blue rectangle and the black line on top represent the cross-section substrate of the tape and the superconducting layer, respectively.}
\end{center}
\end{figure}

We measured the power-law exponent $n$ in a similar way, although we did not make any self-field correction. The reason is that for low magnetic fields, the $n$ exponent is high ($n$ above 30) and for such high $n$ the electro-magnetic response in alternating currents or magnetic fields weakly depends on this parameter. For the same reason, we assume a periodicity of 90 degrees for the angular dependence. Therefore, we enter directly the measured data in the model (with a bi-linear interpolation in $B$ and $\theta$), since an analytical fit is no longer necessary.

\subsection{Coils and AC loss measurement}
\label{s.expcoils}

We constructed four identical pancake coils of 24 turns each with internal and external diameters 60 and 67.8 mm, respectively, as detailed in \cite{pardo12SSTb} (see figure \ref{f.coils}). Afterwards, we pile the pancakes in stacks of 1 to 4 units, with a total height of 4.0, 8.9, 13.1 and 17.6 mm, respectively. Finally, the AC loss was measured by electrical means as follows. The voltage signal is taken from the taps at the terminals. The transport AC current is measured by a Rogowski coil. We also use this voltage (shifted 90 o with respect to the transport current) to compensate the huge inductive component of the measured voltage of the pancake. We set the desired value of this compensation signal by means of a Dewetron DAQP Bridge-B amplifier (sketch in figure \ref{f.setup}).

\begin{figure}[tbp]
\begin{center}
\includegraphics[height=5cm]{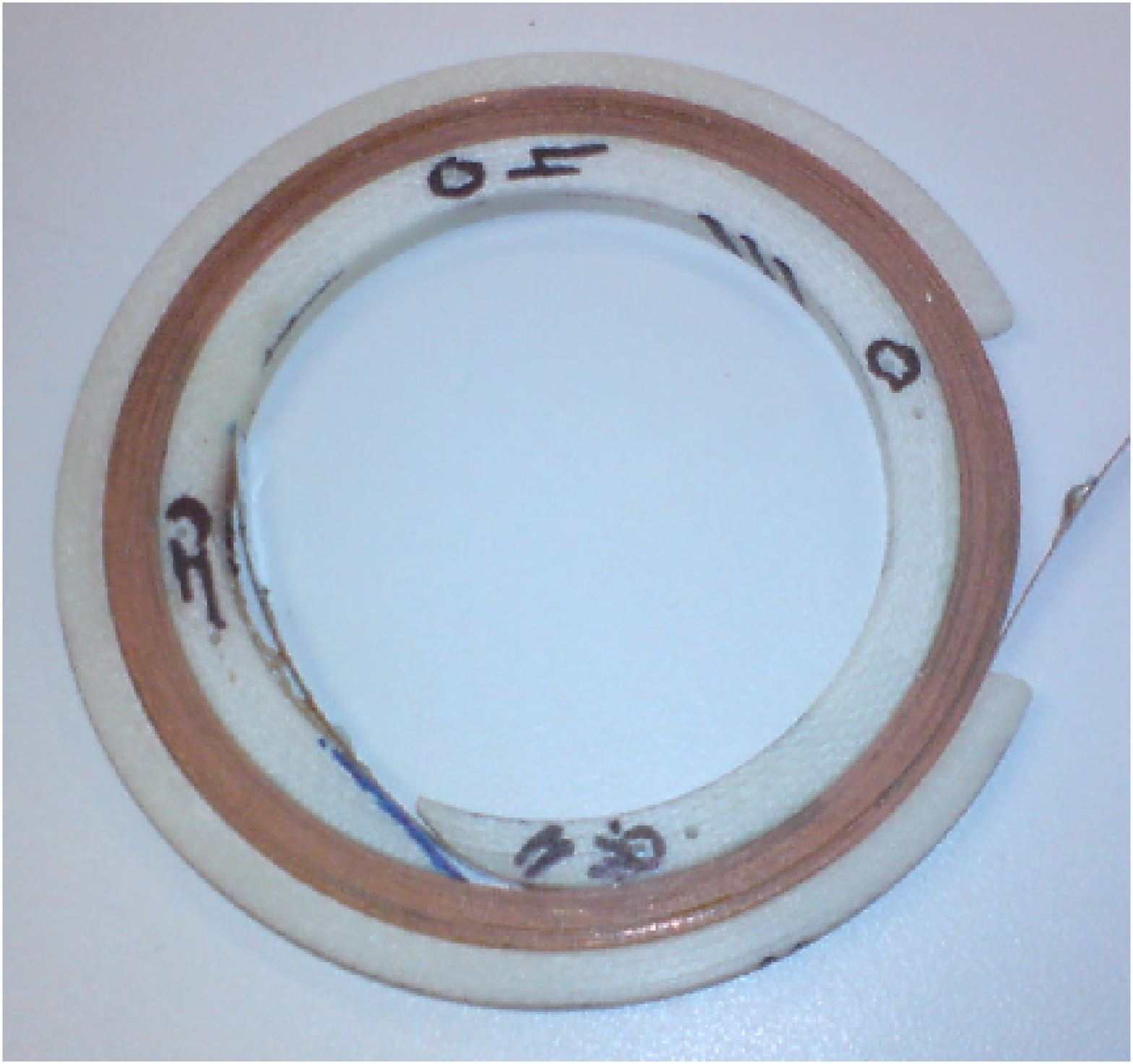}%
\includegraphics[height=5cm]{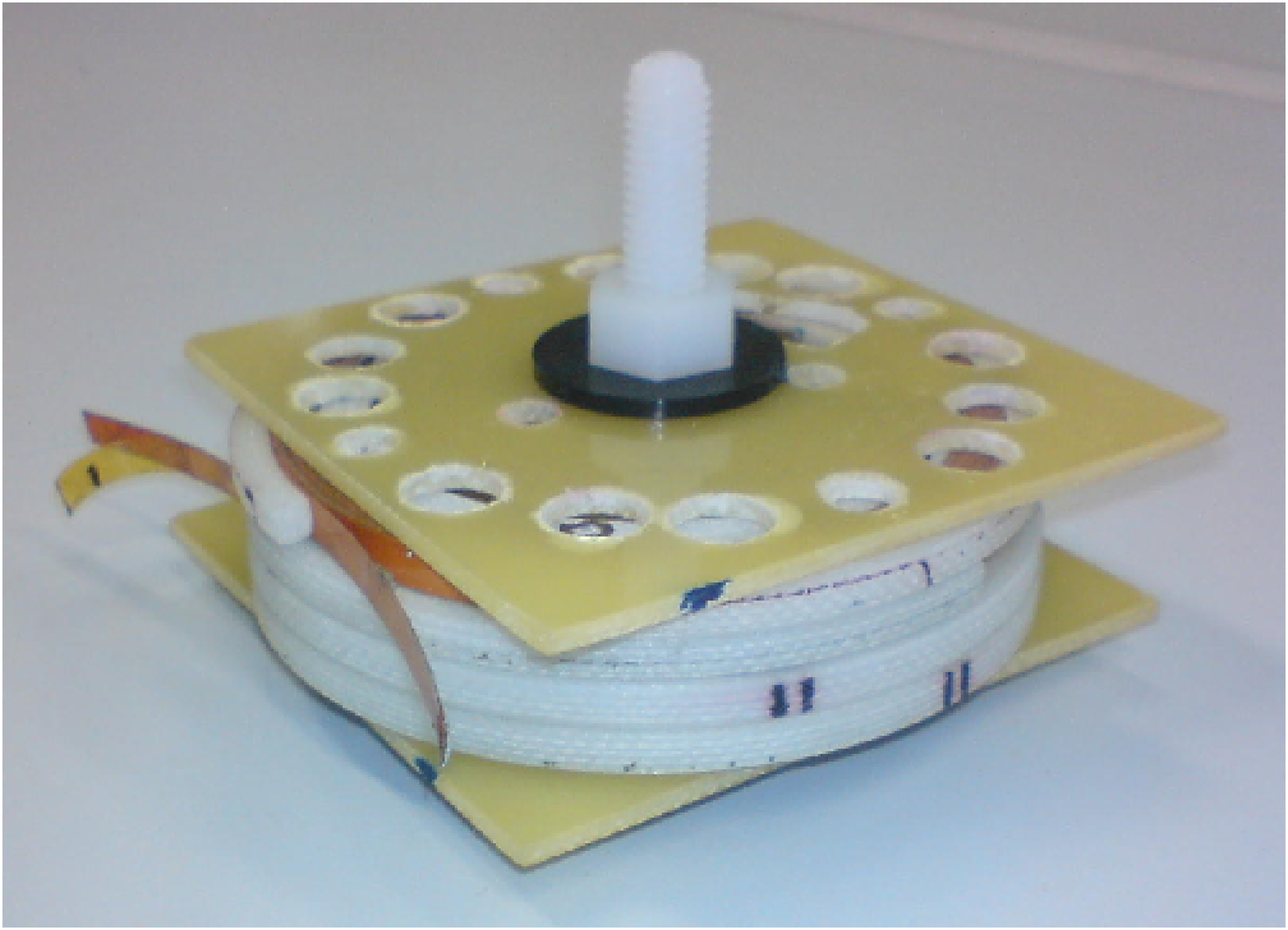}%
\caption{\label{f.coils} Left: one of the pancake coils in the AC loss measurements. Right: stack of 4 pancake coils with mechanical support structure.}
\end{center}
\end{figure}

\begin{figure}[tbp]
\begin{center}
\includegraphics[width=9cm]{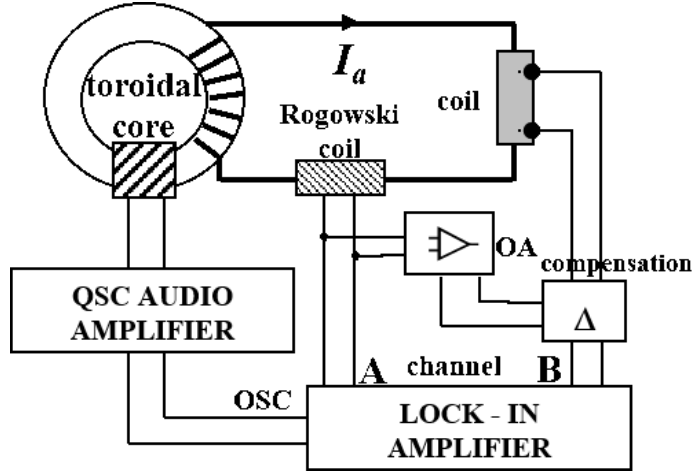}%
\end{center}
\caption{\label{f.setup} Set-up to measure the AC loss in the constructed coils.}
\end{figure}


\section{Benchmarking and comparison with experiments}
\label{s.valid}

This section tests the numerical method by comparing to analytical formulas for thin strips and experiments on stacks of pancake coils, finding a very good agreement for all cases.


\subsection{Single strip}

\begin{figure}[tbp]
\begin{center}
\includegraphics[width=11cm]{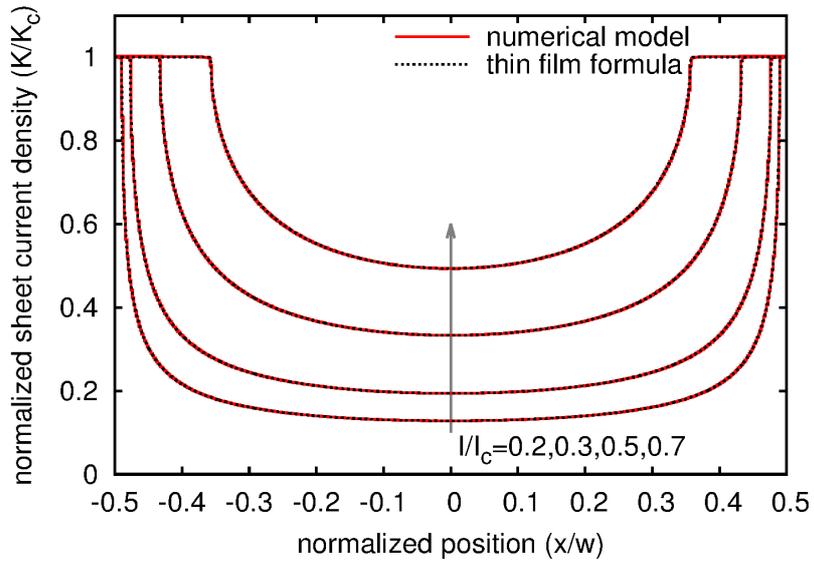}%
\caption{\label{f.Kstrip} The sheet current density $K$ from the numerical model agrees with Norris' thin film formula \cite{norris70JPD}. These results are normalized to the critical sheet current density, $K_c=J_cd$, and the horizontal position $x$ is divided to the tape width $w$. The numerical calculations use a power-law exponent $n=1000$ to describe the critical state model.}\label{f.KNorris}
\end{center}
\end{figure}

In this section, we check the numerical model by comparing the results to analytical formulas for thin strips. In order to compare to our method assuming cylindrical symmetry, we take a single-turn coil with radius much larger than the tape width. In particular, we take an inner radius of 1 m, tape width 4 mm (in the $z$ direction) and thickness 1$\mu$m (in the radial direction).

First, we compare the sheet current density $K$ (current density integrated over the tape thickness) to Norris' formulas for the critical-state model \cite{norris70JPD}
\begin{equation}
K(x)=
\begin{cases}
\frac{2K_c}{\pi}\arctan\sqrt{\frac{(w/2)^2-b^2}{b^2-x^2}} & $for $|x|<b \\
K_c & $for $b<|z|<w/2
\end{cases}
\end{equation}
with $b=(w/2)\sqrt{1-(I/I_m)^2}$ and $K_c=J_cd$, where $d$ is the strip thickness.
The sheet current density from the numerical model for a power-law exponent $n=1000$ coincides to the analytical results (see figure \ref{f.KNorris}). Since the current density agrees with the analytical result, it is not necessary to also compare the AC loss. Actually, checking the current density is more strict than the AC loss, since the current density that produces a given AC loss is not unique. The results in figure \ref{f.KNorris} were computed with 500 elements and a tolerance for $J$ of 0.002 of \% $J_c$. The calculations are for a frequency of 50 Hz, although the results are virtually independent on this parameter.

\begin{figure}[tbp]
\begin{center}
\includegraphics[width=9cm]{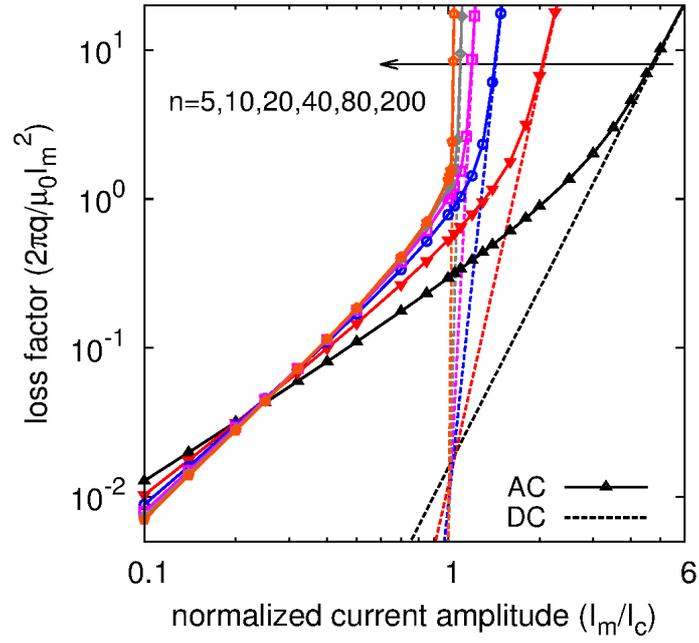}%
\caption{The AC loss for a single strip with constant $J_c$ (solid lines with symbols) approaches to the DC limit at high current amplitudes (dash lines). The vertical axis is the dimensionless loss factor $2\pi Q_l/(\mu_0I_m^2)$, where $I_m$ is the current amplitude and $Q_l$ is the loss per cycle and unit length. The results are for a power-law $E(J)$ relation with different $n$ exponents 5,10,20,40,80,200 (in the arrow direction) and frequency 100 Hz. Symbols (and colors) distinguish lines with different $n$.}\label{f.Gnstrip}
\end{center}
\end{figure}

For low power-law exponents, there do not exist analytical formulas for the current density or the AC loss. Nevertheless, the AC loss should approach to the DC limit at high current amplitudes. This DC loss per cycle and unit tape length is
\begin{equation}
q_{DC}=\frac{S}{T}\oint \dif t E(J_{DC})J_{DC},
\end{equation}
where $S$ is the tape cross-section area and the DC current density is $J_{DC}=|I|/S$. For a sinusoidal excitation $I=I_m\sin(\omega t)$ and a power-law $E(J)$ relation, this DC loss becomes
\begin{equation}
q_{DC}=c(n)E_c \left ( \frac{I_m}{I_c}\right ) ^n I_m
\end{equation}
with
\begin{equation}
c(n)\equiv\frac{2}{\pi}\int_0^{\pi/2} \dif \theta (\sin\theta)^{n+1}.
\end{equation}
For an integer $n$, this function can be evaluated analytically as
\begin{equation}
c(n) =
\begin{cases}
\frac{2}{\pi} \frac{[(n/2)!]^2 2^{n}}{(n+1)!} & $if $n$ is even$\\
\frac{(n+1)!}{\{[(n+1)/2]!\}^2 2^{n+1}} & $if $n$ is odd$.
\end{cases}
\end{equation}
In case that $n$ is non-integer $c(n)$ is calculated numerically. Using the DC loss from the equations above, we found that the computed AC loss for a thin strip approaches to the DC limit for high current amplitudes, which supports the validity of the numerical model (see figure \ref{f.Gnstrip}). In that figure, we plot the loss factor $\Gamma=2\pi q/(\mu_0I_m^2)$, where $q$ is the loss per cycle and tape length, in order to emphasize the differences between curves.

Figure \ref{f.Gnstrip} also shows that for moderate and high $I_m$ ($I_m$ above around $0.3I_c$) the AC loss increases with increasing $n$, while for low current amplitudes the curves follow the opposite trend. The reasons are the following. For high current amplitudes, the AC loss is mostly originated in the region with $J> J_c$; and thence the same $J$ creates lower $E$ for lower $n$, resulting in a decrease of AC loss with decreasing $n$. On the contrary, for low current amplitudes the contribution to the AC loss from the non-critical region ($J<J_c$) becomes important for low $n$. In that case, $E$ decreases with $n$ for a fixed $J$, and thus the AC loss decreases with $n$ until it saturates for high $n$. A similar behaviour has also been observed for round wires \cite{chen05APL,stenvall10SSTa} and multi-filamentary Bi2223 tapes \cite{stavrevthesis}. The $n$ dependence for a fixed current amplitude have also been studied in \cite{sirois08IES}.


\subsection{Comparison with experiments: stack of pancake coils}
\label{s.compexp}

In the following section, we compare the AC loss calculations with measurements for the experimental stacks of pancake coils (see section \ref{s.expcoils} and \cite{pardo12SSTb}), showing a good agreement.

\begin{figure}[tbp]
\begin{center}
\includegraphics[width=9cm]{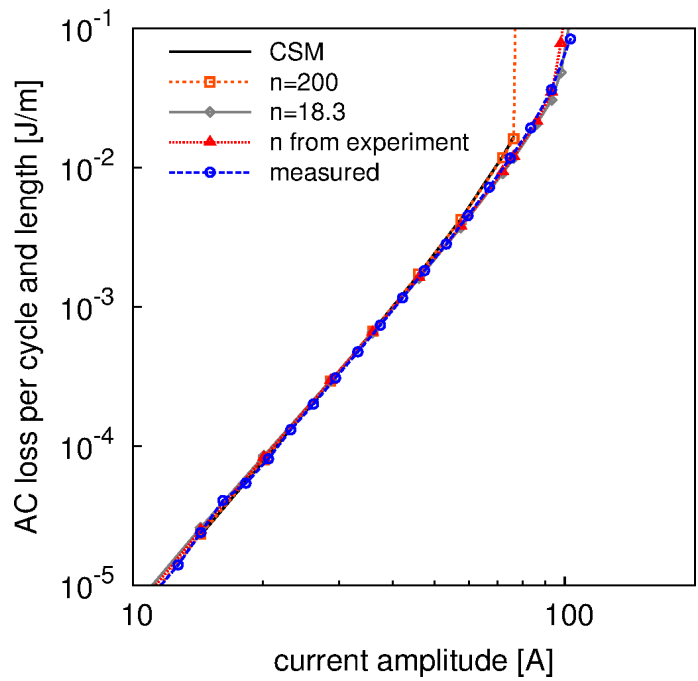}%
\caption{\label{f.Qmeas1x24} Computed AC loss with the Critical State Model (CSM) and power-law $E(J)$ relation with different $n$ exponents compared to the measured one for a single pancake coil (frequency 36 Hz). The curve ``$n$ from experiment" corresponds to the data from figure \ref{f.nBth}. A smooth $E(J)$ relation is necessary to describe the over-critical situation. }
\end{center}
\end{figure}

The AC loss for a single pancake coil from (figure \ref{f.Qmeas1x24}) reveals several features. First, the numerical calculations with $n=200$ coincide with those from \cite{pardo12SSTb} for $n=\infty$ (or the critical state model), supporting again the validity of the method presented here. Second, using a smooth power-law $E(J)$ allows predicting the behaviour for transport currents beyond the critical one, which is not possible for the critical-state model. However, the description presented in this article only allows to calculate situations close beyond the critical current. The limitations are due to the fact that for a high enough current, there will be significant current sharing with the stabilization layers and, in addition, the high dissipation will increase the superconductor temperature, eventually experiencing electro-thermal quench behaviour. The agreement of the model with experiments up to relatively high currents (130\% of $I_c$) can be explained by the relatively low $n$ exponent in the $I_c$-limiting turn ($n\approx 18$), reducing the heating and current sharing effects. Third, using the experimental $n(B,\theta)$ relation of figure \ref{f.nBth} provides practically the same loss results than for the minimum $n$ exponent in the measured range, 18.3. The reason is that the AC loss changes little with small changes of $n$ (see curves in figure \ref{f.Gnstrip} for $n=20$ and 40). In addition, the difference in AC loss becomes larger at high $I_m$, in consistence with the behaviour for a thin strip in figure \ref{f.Gnstrip}. Finally, the computed AC loss for the coil agrees with the measurements for all current amplitudes (see figure \ref{f.Qmeas1x24}). The small discrepancies at the highest amplitudes may be due to experimental error in $J_c$ and $n$, the extraction of $J_c$ from measurements, the assumption that $n$ follows a 90 degrees periodicity, and possible partial current flow in the copper stabilization.

Next, we discuss the AC loss for the stacks from 1 to 4 pancake coils. Figure \ref{f.Qmeas4x24}a presents the calculations for the critical-state model obtained in \cite{pardo12SSTb}, while figure \ref{f.Qmeas4x24}b is for the results of the model in this work with the measured power-law exponent from figure \ref{f.nBth}. For all sets of pancakes, the calculations with smooth $E(J)$ relation agree better with the experiments than for the critical-state model. The agreement is perfect within the measurement error except for the following cases, where there are slight deviations. First, the model over-estimates the AC loss for stacks of 1 and 2 pancakes at very high currents, with the same causes as discussed above for one single pancake. Second, for a very low amplitudes, there is a slight over-estimation of the AC loss, which may be a consequence of avoiding self-field corrections in $n(B,\theta)$. Then, for low magnetic fields, $n$ is actually larger than the one that the model assumes, slightly over-estimating the AC loss. Finally, the computed loss is slightly above the measured one for the whole curve corresponding to 4 pancakes. This could be caused by non-uniformity in the tape length, so that one of the pancakes exposed to the highest AC loss (top and bottom ones) are made of a tape with slightly larger $J_c$. Additionally, there may also be slight errors in the $J_c$ extraction process.

\begin{figure}[tbp]
\begin{center}
\includegraphics[width=9cm]{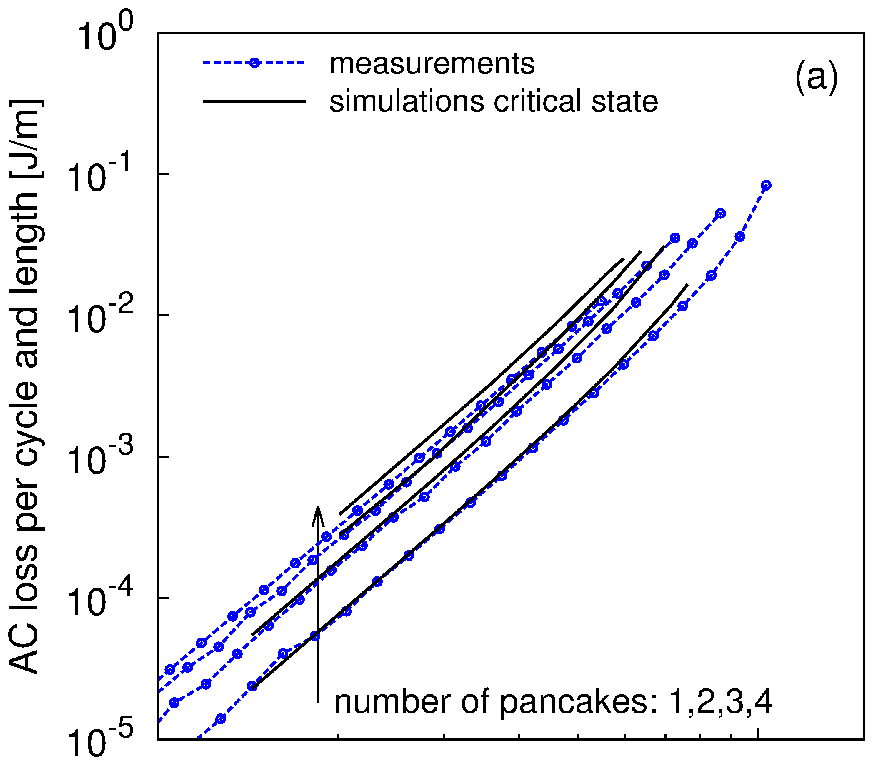} \\%
\includegraphics[width=9cm]{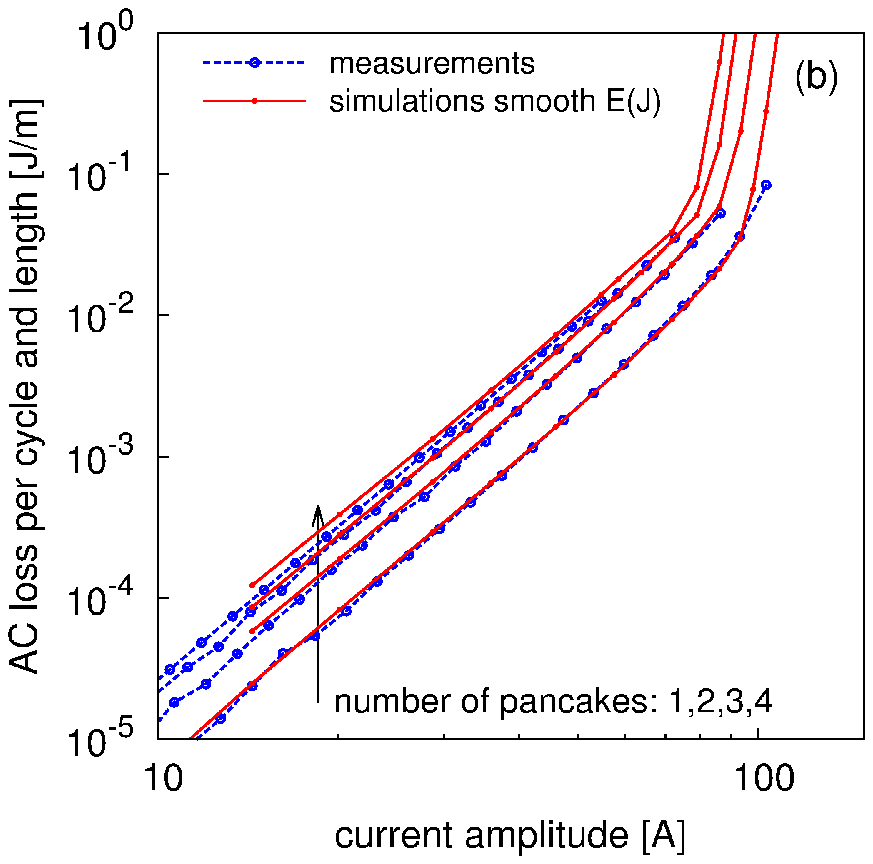}%
\caption{\label{f.Qmeas4x24}  Comparison between measurements and calculations for (a) the critical state model and (b) power-law $E(J)$ relation for the experimental stacks of pancake coils consisting of 1,2,3 and 4 pancakes at 36 Hz frequency. Although the critical-state model agrees with the measurements for mid and low currents, the smooth $E(J)$ relation provides good agreement for all amplitudes, also above the critical current.}
\end{center}
\end{figure}



\section{Magnet-size coils}
\label{s.magnet}

In this section, we apply our numerical model in order to predict several features of an example of magnet-size coil consisting on a stack of pancake coils, such as those in solenoidal SMES \cite{nomura10IES,sander11SST,zhangH13IES,saichi14IES}, high-field magnets \cite{weijers14IES,awaji14IES} or other solenoidal magnets \cite{maeda14IES,daibo13IES,yoonS14IES}. As a generic example of any of these applications, we study a stack of pancake coils consisting on 20 pancakes with 200 turns per pancake (see table \ref{t.magcoil} and figure \ref{f.magsection}). These calculations serve not only to illustrate the model application but also discuss several features for coated conductor coils with many turns. This section presents a purely modelling analysis, confidently based on the comparison of modelling and experimental results from the previous section.

In the following, we present the numerical parameters used in the study (section \ref{s.parnum}), the assumed $J_c(B,\theta)$ relation (section \ref{s.JcBmagnet}), the relaxation effects in $J$ and the generated magnetic field (section \ref{s.rel}), the AC loss (section \ref{s.ACmag}), and the effect of magnetization currents in the magnetic field (section \ref{s.Bqual}).

\begin{figure}[tbp]
\begin{center}
\includegraphics[width=9cm]{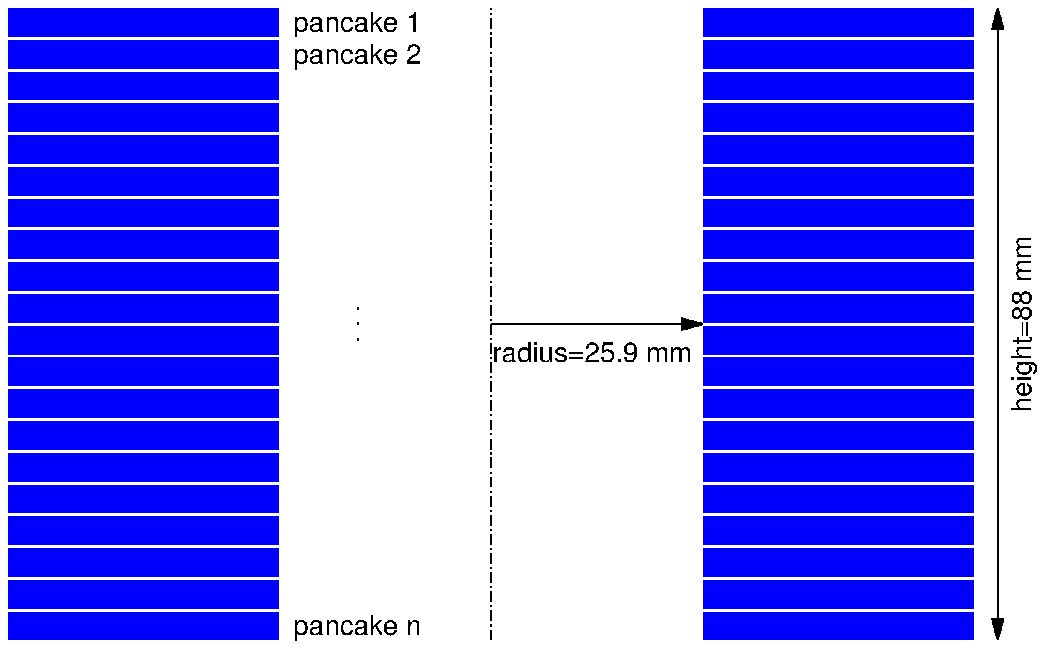}%
\caption{\label{f.magsection} The sketch shows the cross-section of the studied coil, where each rectangle represents a pancake coil. Further details in table \ref{t.magcoil}.}
\end{center}
\end{figure}

\begin{table}
\begin{center}
\footnotesize\rm
\begin{tabular}{ll}
\hline
Number of pancakes & 20 \\
Number of turns per pancake & 200 \\
Inner radius & 29.5 mm \\
Outer radius & 67.2 mm \\
Total heigh & 88.0 mm \\
Tape width & 4.0 mm \\
\hline
\end{tabular}
\end{center}
\caption{Geometrical parameters of the modelled example of magnet-size coil.}\label{t.magcoil}
\end{table}

\subsection{Numerical parameters}
\label{s.parnum}

In order to simplify the computations, we take the continuous approximation, that is we approximate each pancake coil as a continuous object with the same engineering current density as the original one \cite{prigozhin11SST,neighbour,zermeno13JAP}. As shown in \cite{neighbour}, this approximation introduces negligible errors, providing a slight under-estimation in the AC loss at very low current amplitudes. With this approximation, we divide the coil radial thickness into 20 equivalent turns of identical cross-section and no separation between them, which transport 10 times the current of one turn in the original 200-turn pancake coil. We also use 50 elements across the tape width and a tolerance of $J$ between 0.008 and 0.002 \% of $J_c(B=0)$, being the lowest values for the lowest current amplitudes or for magnetic relaxation calculations.


\subsection{Assumed magnetic field dependence on $J_c$ and coil critical current}
\label{s.JcBmagnet}

\begin{figure}[tbp]
\begin{center}
\includegraphics[width=9cm]{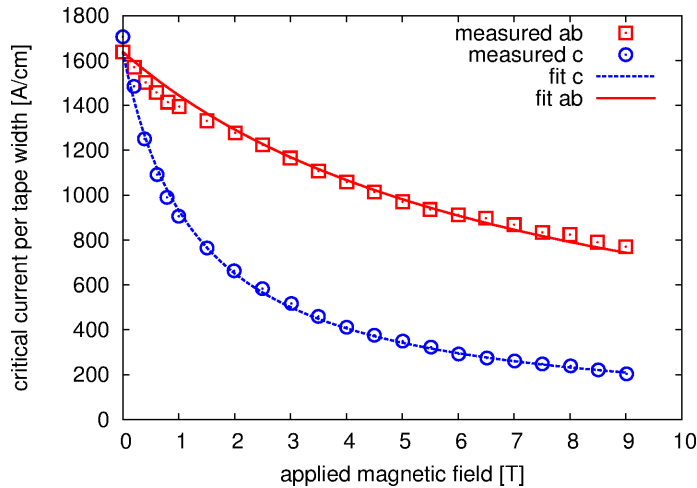}%
\caption{\label{f.IcBselva} Magnetic field dependence of $J_c$ used in the model for the coil in figure \ref{f.magsection}. Symbols are measurements at 50 K from Selvanickam {\it et al.} in \cite{selvamanickam12SST} and lines are analytical fits from equations (\ref{f.Ji}) and (\ref{f.Jl}).}
\end{center}
\end{figure}

In order to generate magnetic fields of considerable magnitude, operating temperatures well below 77 K are necessary, due to the severe reduction or the critical current above 1 T at this temperature \cite{daibo13IES,yoonS14IES}. Setting a goal of 7 T of generated magnetic field in the bore, our coil requires a critical current of 190 A, which cannot be achieved at 77 K based on current material performance. Therefore, the experimental $J_c$ data for 77 K in section \ref{s.Jcn} is not useful for this case. Instead, we use the data from \cite{selvamanickam12SST} for 50 K. For simplicity, we took the measured data in that article for applied magnetic fields in the perpendicular and parallel directions and fit the magnetic field dependence with a Kim-like function \cite{kim62PRL}. In addition, we simplified the angular dependence as an elliptical function. Therefore, the assumed magnetic field dependence of $J_c$ is
\begin{equation}
\label{JcBmag}
J_c(B,\theta)=\frac{J_{c0}}{1+\frac{Bf(\theta)}{B_0}}
\end{equation}
with
\begin{equation}
\label{Jcthmag}
f(\theta)=\sqrt{u^2\cos^2\theta+\sin^2\theta},
\end{equation}
where $J_{c0}$, $B_0$, $u$ are constant parameters. The dependence from the equation above (actually $J_cd$) fits well to the experimental data from \cite{selvamanickam12SST} for parallel and perpendicular applied magnetic field for $J_{c0}=1.405\times 10^{11}$ A/m$^2$, $B_0=7.47$ T, $u=5.66$, where we assumed a superconducting layer thickness of $d=1.4$ $\mu$m (see figure \ref{f.IcBselva}). Note that taking $J_c$ directly from $I_c$ measurements, as done for equations (\ref{JcBmag}) and (\ref{Jcthmag}), neglects the self-field effects in the $I_c$ measurements. However, the error in the taken $J_c$ is negligible for magnet-size coils because the local magnetic fields are high. Additionally, although the real angular dependence is more complex than the assumed elliptical type, the obtained results with the above $J_c(B,\theta)$ provide the main features for magnet-size coils. As shown in section \ref{s.compexp}, the numerical model can use a more complex angular dependence, if provided for several applied magnetic fields.

With these parameters, we calculated the coil critical current as that of the weakest turn (as defined in previous works \cite{coatedIc,pitel13SST}), with a result of $I_{c,{\rm coil}}=194$ A. In order to be sure that this value corresponds to the DC limit, we have computed the critical current by increasing the current in a quarter sinusoidal cycle of $10^{-14}$ Hz. The critical current is determined by evaluating the voltage per unit length in all the turns (see section \ref{s.vphi}) and using a voltage criterion of 1 $\mu$V/cm.

In this article, we arbitrarily chose a power-law exponent of 20, although the method allows to calculate any exponent without significant variation in the computing time.


\subsection{Relaxation effects}
\label{s.rel}

Next, we study the relaxation effects after energizing the coil and keeping the current constant for a certain time. In particular, we analyze the case of increasing the current up to 162 A following a quarter sinusoidal cycle of 0.1 Hz (charging curve of 2.5 s) and afterwards keeping the current constant for one hour. The calculations for this case use a time step that increases exponentially.

\begin{figure}[tbp]
\begin{center}
\includegraphics[width=9cm]{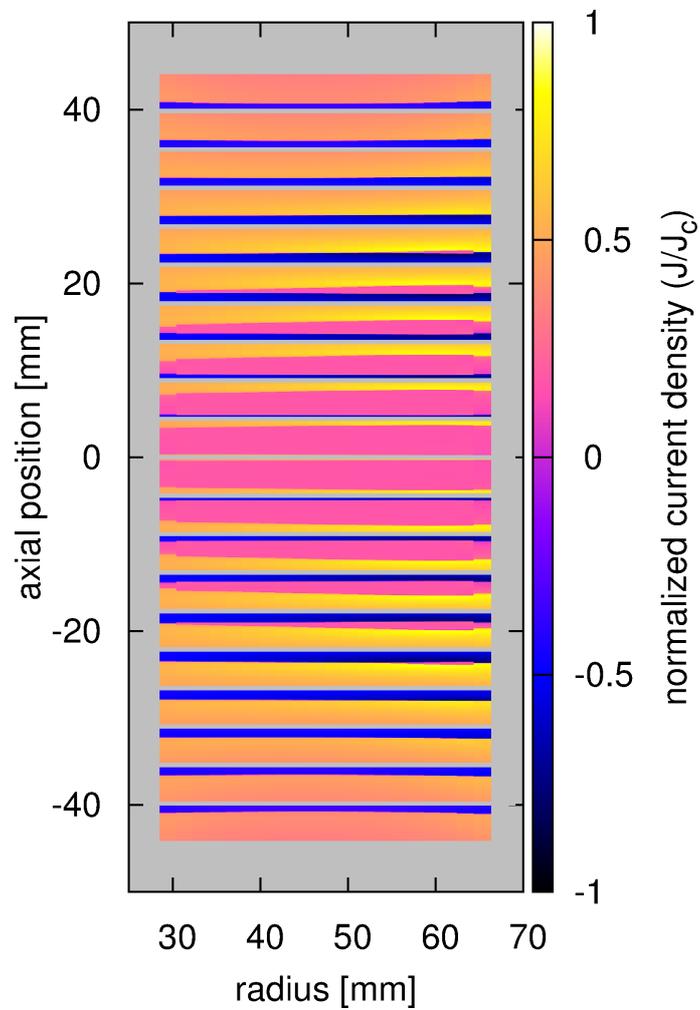}%
\caption{\label{f.Ji} Current density in the coil in figure \ref{f.magsection} at the end of the charging curve, consisting on a quarter sinusoidal cycle of 162 A peak and 0.1 Hz frequency. Negative current density evidences magnetization currents.}
\end{center}
\end{figure}

\begin{figure}[tbp]
\begin{center}
\includegraphics[width=9cm]{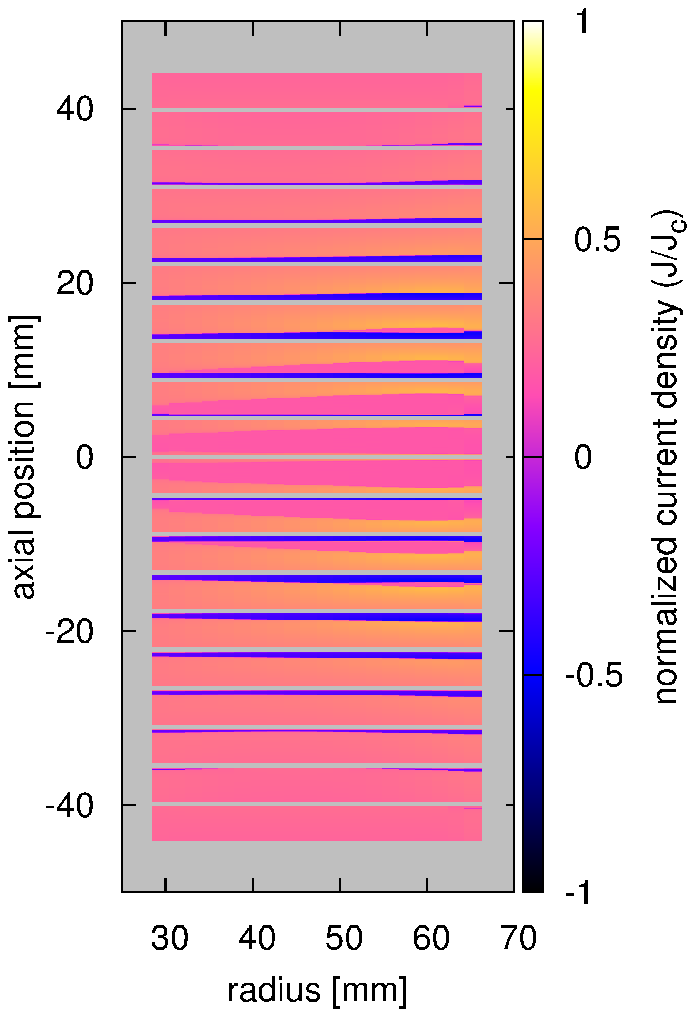}%
\caption{\label{f.Jl} Current density in the coil from figure \ref{f.magsection} for 1 hour relaxation after the charging curve (situation at the end of charging curve in figure \ref{f.Ji}). Magnetization currents are strongly suppressed.}
\end{center}
\end{figure}

At the end of the charging curve, the presence of current density with opposite sign to the transport current is evidence of important magnetization currents (figure \ref{f.Ji}). The 4 top and bottom pancakes are saturated with magnetization currents. After one hour of relaxation, the magnetization currents are strongly suppressed, disappearing from the top and bottom pancakes (see figure \ref{f.Jl}) and the current density becomes more uniform in all pancakes.

In more detail, the current density at the end of the charging curve (figure \ref{f.Ji}) presents the same qualitative features as for the critical-state model with constant $J_c$ \cite{neighbour}. Apart from the fact that the pancakes closest to the top and bottom are saturated with magnetization currents, there appears a sub-critical zone in the rest of the pancakes, where $J$ is uniform with roughly the same value for all the pancakes and proportional to the net current in the coil. The main additional feature appearing in the calculations of the present work is that in the critical region (region with $J\ge J_c$) the current density decreases from the edge of the sub-critical zone or border between positive and negative $J$ with approaching the top and bottom edge of each pancake. The cause of this effect is the increase of the radial magnetic field, since it vanishes at the sub-critical zone \cite{pardo12SSTb} and becomes minimum at the border between the positive and negative $J$ due to the magnetic field created by the magnetization currents.

\begin{figure}[tbp]
\begin{center}
\includegraphics[width=9cm]{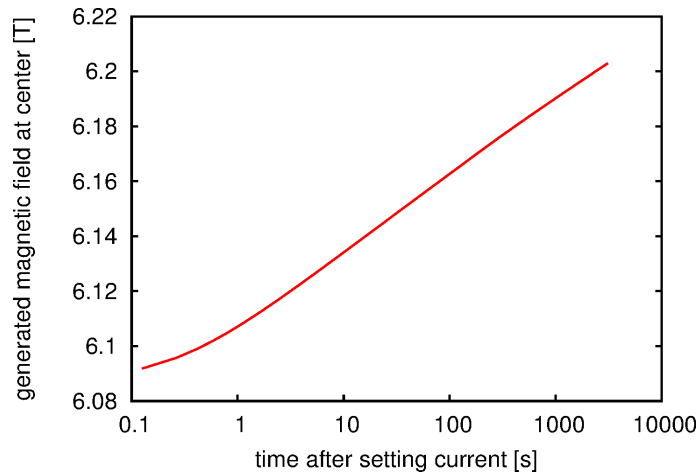}%
\caption{\label{f.Brel} The relaxation of magnetization currents causes that the generated magnetic field at the bore center increases with time after setting the current to a certain constant value (162 A).}
\end{center}
\end{figure}

The relaxation of current density has an important effect on the generated magnetic field at the bore center (figure \ref{f.Brel}). After one hour relaxation, the generated magnetic field increases by around 100 mT on a background magnetic field of approximately 6 T, representing roughly 1.7 \% increase. This increase is relatively high for magnets and may not be suitable for certain applications, such as Nuclear Magnetic Resonance or accelerator magnets. However, coil optimization can reduce the impact of the relaxation effect. For superconducting tapes with higher power-law exponents, this increase in the generated magnetic field will require higher relaxation times but it will still be present.


\subsection{AC loss}
\label{s.ACmag}

This section discusses the AC loss due to alternating transport currents at 0.1 Hz.

\begin{figure}[tbp]
\begin{center}
\includegraphics[width=9cm]{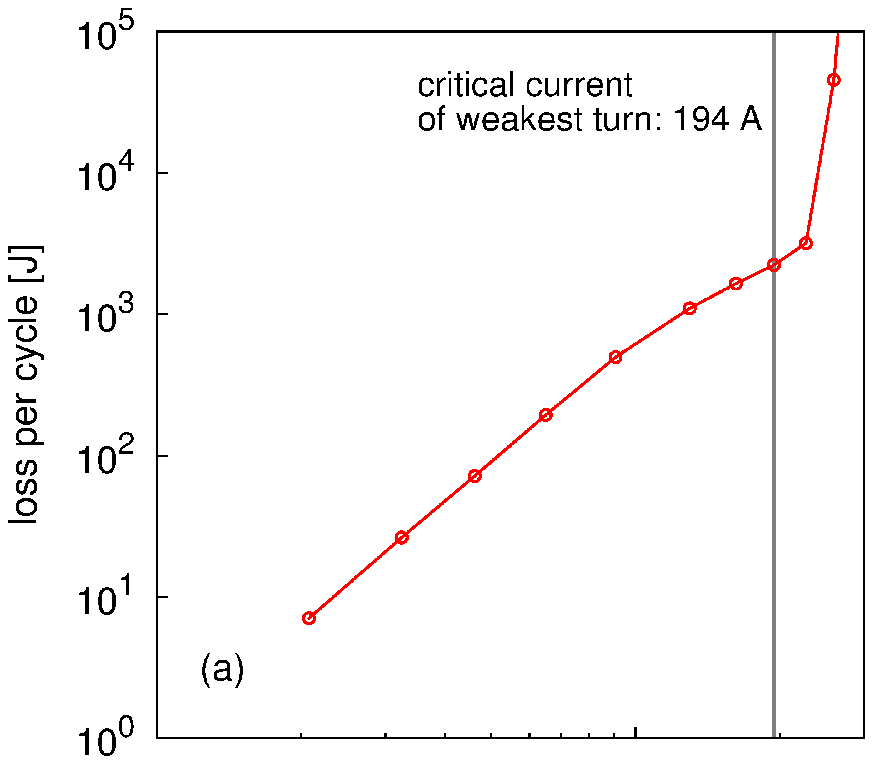} \\%
\includegraphics[width=9cm]{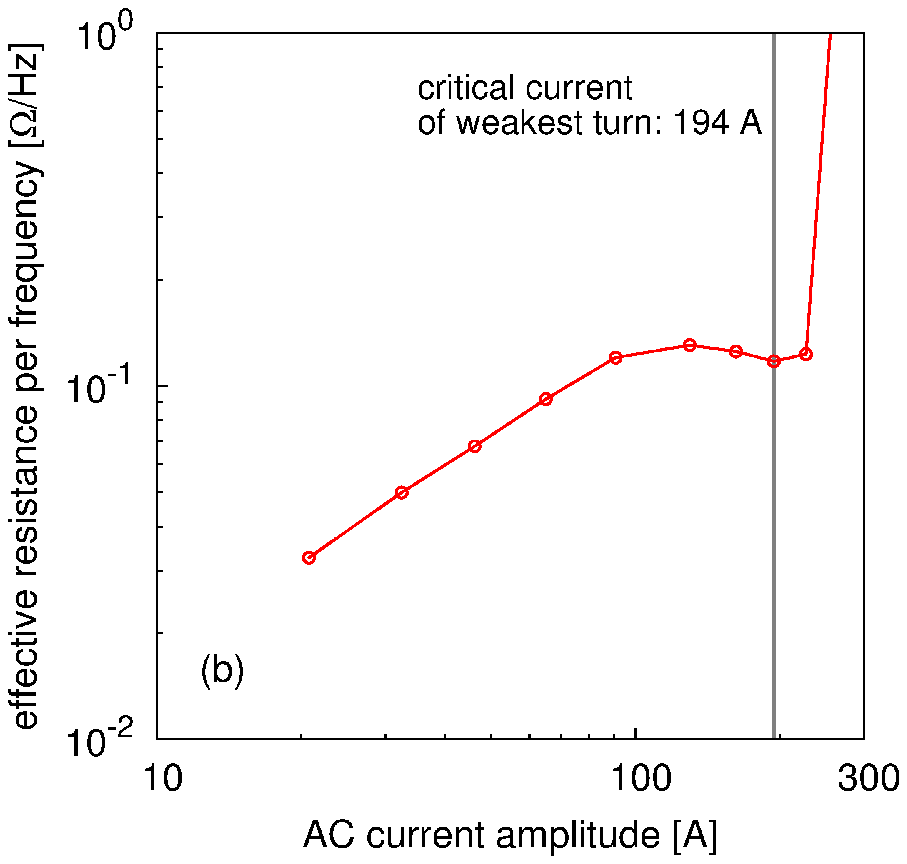}%
\caption{\label{f.Q} (a) Calculated AC loss per cycle $Q$ and (b) computed effective resistance per frequency $R_{\rm eff}=2Q/(fI_m^2)$. For both cases, the AC frequency is 0.1 Hz. The peak in $R_{\rm eff}$ evidences magnetization loss.}
\end{center}
\end{figure}

The AC loss increases with increasing the AC current (see figure \ref{f.Q}a), presenting the following features. At low amplitudes, the loss curve in logarithmic scale presents a slope of around 3; with growing the AC current, the slope decreases down to roughly 1.7; finally, at very high AC currents the slope sharply increases to a value between 20 and 30. The slope of around 3 at low amplitudes and its following decrease can be explained by Bean's slab model for magnetization loss \cite{bean64RMP,goldfarbSpin,clemSpin,reviewac}, since a pancake coil with many turns under a radial applied magnetic field roughly behaves as a slab. In that case, the AC loss is proportional to $H_m^3$, where $H_m$ is the applied magnetic field amplitude, until the slab penetrates. Beyond the $H_m$ where this occurs, the AC loss gradually becomes proportional to $H_m$. Since in a pancake coil of our winding, $H_m$ generated by the other coils is proportional to $I_m$, the loss curve as a function of $I_m$ should follow the same dependence. The fact that in our coil the slope does not decrease as much as 1 is caused by, first, the contribution from the transport loss and dynamic magneto-resistance (as seen for a single tape in \cite{HacIacinphase}) and, second, the onset of the non-linear resistive loss for transport currents below the critical value. Actually, the latter contribution is the responsible of the sharp slope rise at very high currents. A slope higher than the $n$ power-law exponent of 20 at the over-critical situation is caused by the decrease of $J_c$ with the increase of the magnetic field when increasing $I_m$. Note that this contribution to the AC loss is not apparent for currents just above the critical values, requiring $I_m\approx 1.17 I_c$ in our case. The reason is that, according to our definition of coil critical current, at the coil $I_c$ only one turn is above the local $I_c$, and thence the resistive loss contribution to the total AC loss is small.

The AC loss behaviour is more evident when represented as the quantity $R_{\rm eff}=2Q/I_m^2$ (see figure \ref{f.Q}b). This quantity has the interpretation of an effective resistance per unit frequency, since the power loss in a device of resistance $R$ is $P=\frac{1}{2}RI_m^2$. With this representation, the saturation of the magnetization loss appears as a peak.


\subsection{Magnetic field distortion due to magnetization currents}
\label{s.Bqual}

In this section, we discuss the effect of the generated magnetic field of the magnetization currents and the distortion that they create. For this purpose, we consider alternating currents of 0.1 Hz frequency.

\begin{figure}[tbp]
\begin{center}
\includegraphics[width=9cm]{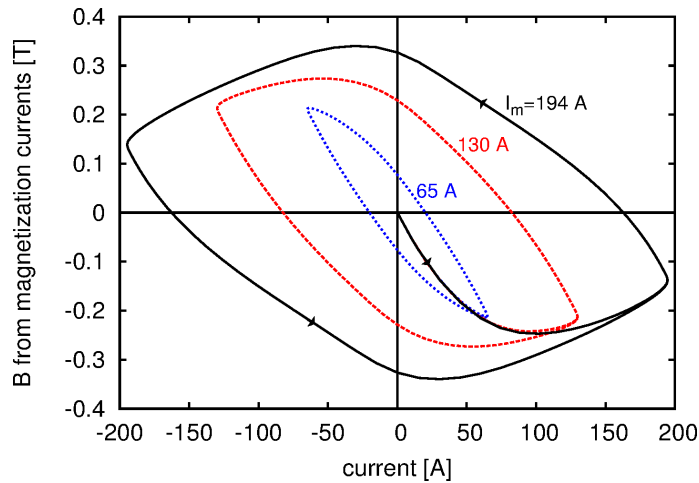}%
\caption{\label{f.Bcencyc} Magnetic field at the bore center generated by magnetization currents (total magnetic field minus magnetic field assuming uniform current in the tape) for alternating current of 0.1 Hz. The arrows at the curves for the largest current amplitude represent the current sequence starting from the zero-field-cool situation.}
\end{center}
\end{figure}

First, we analyze the magnetic field at the bore center due to magnetization currents, $B_{\rm c,mag}$, in figure \ref{f.Bcencyc}. This contribution to the generated magnetic field presents a hysteresis cycle with a previous initial curve. With increasing the current, the absolute value of $B_{\rm c, mag}$ at the initial curve first increases, presents a peak, and then decreases. The cause of the initial rise is the creation of magnetization currents, while the reason of the decrease at higher currents is both the decrease of $J_c$ due to the higher local magnetic field and the depletion of magnetization currents due to the increase of transport current. It is important to notice that the remanence is relatively high, up to 330 mT, being 4.5 \% of the maximum generated field at the critical current, 7.3 T.

\begin{figure}[tbp]
\begin{center}
\includegraphics[width=9cm]{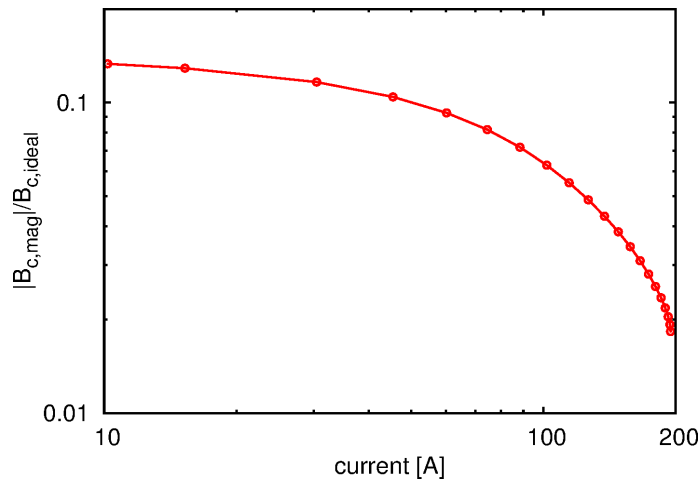}%
\caption{\label{f.Bcenin} Contribution of the magnetization currents to the generated magnetic field in the bore center, $B_{c,{\rm mag}}$, relative to the generated magnetic field if magnetization currents are not present, $B_{c,{\rm ideal}}$. The plotted results are for the initial curve.}
\end{center}
\end{figure}

An important parameter for magnet technology is the magnetic field distortion. That is, the relative error of the generated magnetic field compared to the design value. Often, the design value is taken as the magnetic field created by the winding, ignoring magnetization currents \cite{amemiya08SST,ahnMC09IES,yanagisawa11IES,amemiya12PhC} ($B_{\rm ideal})$, and therefore the magnetic field distortion is $|B_{\rm c,mag}|/B_{\rm ideal}$. For our example, this magnetic field distortion at the bore center and the initial curve ranges between 0.13 and 0.018, decreasing with the current (figure \ref{f.Bcenin}). These values are very high for NMR and accelerator magnets \cite{amemiya12PhC,zhangM14SST}, although a decrease of this quality factor could be obtained by optimizing the winding geometry. However, magnets with small bores are likely to present low quality factors, since the magnetic field created by magnetization currents increases with decreasing the distance from the winding. Another way to reduce the magnetic field distortion could be by taking magnetization currents into account in the magnet design.


\section{Conclusion}
\label{s.conc}

Summarizing, this article has presented a method to numerically calculate the electromagnetic properties of superconductors described by any ${\bf E}({\bf J})$ relation under slowly varying magnetic fields. For this purpose, we have obtained a variational principle in the ${\bf J}-\phi$ formulation that reduces the problem to the sample volume, avoiding unnecessary elements in the air; and thence speeding up the computations. Although this formulation is valid for any 3-dimensional shape, the results in this article are for coils with cylindrical symmetry. For this case, we have presented the details of the numerical method to find the current density and other electromagnetic quantities by minimizing the functional. Afterwards, we have satisfactorily tested the method by comparing to thin-strip formulas and experiments for stacks of pancake coils. Finally, we have applied the method to calculate the AC loss, relaxation effects and magnetic field quality of a magnet-size coil made of 20 pancakes and 200 turns per pancake.

In particular, we have found that our modelling results coincide with the formulas for thin strips. In addition, the AC loss agrees very well with the measurements of a stack of a few pancake coils and, thanks to the smooth $E(J)$ relation, the loss (and other electromagnetic quantities) can also be predicted for over-critical situations. For the magnet-size coil, we have seen that for a power-law exponent of 20, the magnetization currents are substantially suppressed after 1 hour relaxation, appreciably increasing the generated magnetic field. For higher power-law exponents, the same kind of relaxation will occur but with higher relaxation times. Magnetization currents under cyclic input current are also important, decreasing the magnetic field quality. As a consequence, predicting magnetization currents in coated conductors is necessary for magnet design at least for magnets with small bore or strict specifications regarding the quality factor, such as NMR or accelerator magnets.

In conclusion, the modelling tool presented in this article satisfies the requirements to predict the electromagnetic behavior of windings from a few turns to magnet-size coils, as well as other multi-tape arrangements. In addition, the presented variational principle is promising for computationally demanding 3-dimensional problems.


\section*{Acknowledgements}

We acknowledge L. Prigozhin and M. Kapolka for valuable discussions. The authors acknowledge the use of resources provided by the SIVVP project (ERDF, ITMS 26230120002). The research leading to these results has received funding from the European Union Seventh Framework Programme [FP7/2007-2013] (grant NMP-LA-2012-280432) and the Structural Funds of EU (grant ITMS 26240220088)(0.5).


\appendix

\section{Euler-Lagrange equations of a functional}
\label{s.EL}

Here, we summarize the Euler-Lagrange equations for general functionals, for the reader's convenience.

Given a certain functional, depending on a set of variables ${\bf x}=(x_1,x_2,\dots,x_n)$, a set of functions of those variables $a_i({\bf x})$ (with $i\in{1,2,\dots,m}$) and their partial derivatives $\partial_ja_i({\bf x})$, where $\partial_j\equiv\partial/\partial x_j$, such that
\begin{equation}
\label{funcfirst}
L=\int_{V_n}{\rm d}^n{\bf x} f({\bf x},a_i({\bf x}),\partial_ja_i({\bf x})),
\end{equation}
where $V_n$ is any $n$-dimensional volume;
finding an extreme (maximum, minimum or saddle) of that functional regarding variations of the functions $a_i({\bf x})$ is equivalent to solving the following $m$ differential equations (see page 192 of \cite{courant})
\begin{equation}
\label{ELgen}
\frac{\partial f}{\partial a_i}-\sum_{j=1}^n \partial_j \left [ \frac{\partial f}{\partial(\partial_ja_i)} \right ]=0.
\end{equation}
These equations are known as the Euler-Lagrange equations of the functional $L$.

For functionals that also contain second partial derivatives of the functions, $\partial_j\partial_ka_i({\bf x})$, such as
\begin{equation}
\label{funcsecond}
L=\int_{V_n}{\rm d}^n{\bf x} f({\bf x},a_i({\bf x}),\partial_ja_i({\bf x}),\partial_j\partial_ka_i({\bf x})),
\end{equation}
its Euler-Lagrange equations are (see page 192 of \cite{courant})
\begin{equation}
\frac{\partial f}{\partial a_i}-\sum_j\partial_j\left [ \frac{\partial f}{\partial(\partial_ja_i)} \right]+
\sum_{jk} \partial_j\partial_k \left [ \frac{\partial f}{\partial(\partial_j\partial_ka_i)} \right ]=0.
\end{equation}


\section{On the minimum of the functional}
\label{s.convex}

In this appendix, we show that the extreme of the functional in the $\bf J$-$\phi$ formulation of equation (\ref{fDA}) is a minimum at least when $\nabla \phi$ is not directly involved in the extreme-finding process, such as when it is created by an external source like in coils.

Next, we assume a discretization of the system in volumes of constant $\bf J$. The limit when the value of all elements approach to 0 corresponds to the general continuous case. Then, ${\bf J}({\bf r})$ is represented by an array of variables $\{J_{is}\}$, where $i\in[1,N]$ and $s\in\{x,y,z\}$ label the element and $\bf J$ vector component, respectively, and $N$ is the number of elements. Then, the functional is
\begin{equation}
L\approx\frac{1}{2\Delta t}\sum_{ij}\sum_{s}V_iV_j\Delta J_{is}\Delta J_{js} M_{ij}+\sum_i V_i U({\bf J}_i)+\sum_i \partial_s\phi_i J_{is} V_i  ,
\end{equation}
where $V_i$ and $V_j$ are the volumes of elements $i$ and $j$, respectively, and $M_{ij}$ is defined as $M_{ij}\equiv [\mu_0/(4\pi V_iV_j)]\int_{V_i}{\rm d}V\int_{V_j}{\rm d}V'(1/|{\bf r}-{\bf r'}|)$. The Hassian matrix of this discretized functional, with elements $H_{ijsl}$, is
\begin{equation}
\label{Hijsl}
H_{ijsl}\equiv\frac{\partial L}{\partial \Delta J_{is}\partial\Delta J_{jl}}=V_iV_jM_{ij}\delta_{sl}/\Delta t+V_i\delta_{ij}\rho_{sl}({\bf J}_i),
\end{equation}
where $\rho_{sl}=\partial E_s/\partial J_l$ is the differential resistivity matrix. The functional presents a minimum when the Hassian is positive definite; that is, $\sum_{ij}\sum_{sl}J'_{is}J'_{jl}H_{ijsl}>0$ for any $\{J'_{is}\}$. From (\ref{Hijsl}), this sum is
\begin{equation}
\sum_{ij}\sum_{sl}J'_{is}J'_{jl}H_{ijsl}=\frac{1}{\Delta t}\sum_{ij}\sum_{s}V_iV_jM_{ij}J'_{is}J'_{js}+\sum_i\sum_{ls}V_i\rho_{sl}({\bf J}_i)J'_{is}J'_{il}.
\end{equation}
The first term on the right is positive because it is proportional to the self-interaction energy of the currents, which is always positive. The second term is also positive because $\rho_{sl}({\bf J}_i)$ is positive definite for any physical ${\bf E}({\bf J})$ relation, due to thermodynamical reasons. Therefore, the Hassian of the functional is definite positive, and it presents a minimum.


\section{Calculation of the interaction matrices}
\label{s.intmat}

This appendix presents details on the numerical method to calculate the matrix elements $C_{ij}$ and ${\bf b}_{ij}$ in equations (\ref{cij}) and (\ref{bij}). 

First of all, we provide the expressions of the vector potential and magnetic field generated by a circular loop, $a_{\rm loop}$ and ${\bf b}_{\rm loop}$ in SI, for the reader's convenience (formulas is Gauss units in p. 112 of \cite{landau})
\begin{eqnarray}
a_{\rm loop}(r,r',z-z') & = & \frac{\mu_0}{\pi k_e}\left ( {\frac{r'}{r}} \right )^{\frac{1}{2}} \left [ { F_e(k_e)(1-k_e^2/2)-E_e(k_e) } \right ] \label{aloop} \\
b_{r,{\rm loop}}(r,r',z-z') & = & \frac{\mu_0}{2\pi}\frac{z-z'}{\sqrt{(r'+r)^2+(z-z')^2}}  \nonumber \\
&& \cdot\left [ { -F_e(k_e)+E_e(k_e)\frac{r'^2+r^2+(z-z')^2}{(r-r')^2+(z-z')^2} } \right ] \label{brloop} \\
b_{z,{\rm loop}}(r,r',z-z') & = & \frac{\mu_0}{2\pi}\frac{1}{\sqrt{(r'+r)^2+(z-z')^2}}  \nonumber \\
&& \cdot\left [ { F_e(k_e)+E_e(k_e)\frac{r'^2-r^2-(z-z')^2}{(r-r')^2+(z-z')^2} } \right ] \label{bzloop}
\end{eqnarray}
with
\begin{equation}
\label{k}
k_e=2\sqrt{\frac{rr'}{(r+r')^2+(z-z')^2}} ,
\end{equation}
where $b_{r,{\rm loop}}$ and $b_{z,{\rm loop}}$ are the $r$ and $z$ components of ${\bf b}_{\rm loop}$, respectively, $(r,z)$ and $(r',z')$ are the coordinates of the observation point and the loop position, respectively, and $F_e$ and $K_e$ are the complete elliptic integrals of the first and second kind, respectively.

We numerically calculate ${\bf b}_{ij}$ as follows. First, we define the magnetic field generated by one element at any point $(r,z)$ per unit current in that element, being
\begin{equation}
{\bf b}_j(r,z)=\frac{1}{S_j}\int_{S_j} \dif s' {\bf b}_{\rm loop}(r,r',z-z'),
\end{equation}
where ${\bf b}_{\rm loop}$ is given by equations (\ref{brloop})-(\ref{k}). Next, we numerically evaluate ${\bf b}_j({\bf r})$ by dividing element $j$ into sub-elements
\begin{equation}
{\bf b}_j(r,z)\approx \frac{1}{S_j}\sum_{p=1}^{n_r}\sum_{q=1}^{n_z} d_rd_z{\bf b}_{\rm loop}(r,r_p,z-z_q),
\end{equation}
where $r_p=r_j-{\Delta r}/{2}+d_r ( { p-{1}/{2} } )$, $z_q=z_j-{\Delta z}/{2}+d_z ( { q-{1}/{2} } )$, $d_r=\Delta r/n_r$ and $d_z=\Delta z/n_z$, being $(r_j,z_j)$ the center of element $j$ cross-section and $\Delta r$, $\Delta z$ its radial and axial width, respectively. The number of sub-elements in the radial and axial directions, $n_r$ and $n_z$, are determined as follows. We fix $n_z$ to a certain value and obtain $n_r$ such that the sub-element cross-section is as square as possible; $n_r={\rm int}[(\Delta r/\Delta z)n_z+1/2]$, where ${\rm int}(x)$ is the integer part of $x$, and we set a minimum value $n_r=1$. Actually, we have just assumed that $\Delta z>\Delta r$. Otherwise, the determination of $n_z$ and $n_r$ are done accordingly by fixing first $n_r$ and calculating its corresponding $n_z$. In order to achieve a value with a certain given tolerance for a given component of ${\bf b}_j$, for instance the radial component $b_{r,j}$, we calculate first $b_{r,j}$ with $n_z=1$ (if $\Delta z>\Delta r$), re-calculate $b_{r,j}$ after duplicating the value of $n_z$ and repeat the process until the difference in ${\bf b}_j$ between two consecutive values is below a certain relative tolerance. Afterwards, we do the same process for the other component, $b_{z,j}$. In this way, if the observation point is very far away from the element center, we may require only two sub-elements in order to achieve the desired tolerance. We have found that for the computations in this article, decreasing the tolerance below 0.01 \% does not have any influence on the results. Once we are able to calculate ${\bf b}_j({\bf r})$ for any point $\bf r$, we can compute its average in any $i$ cross-section, ${\bf b}_{ij}$, in a similar way. Chiefly, 
\begin{equation}
{\bf b}_{ij}\approx \frac{1}{S_i}\sum_{p'=1}^{n_r'}\sum_{q'=1}^{n_z'} d_r'd_z'{\bf b}_j(r_p',z_q'),
\end{equation}
where $r_p'$ and $z_q'$ are defined analogously as $r_p$ and $r_q$ above. That is, $r_p'=r_i-\Delta r/2+d_r'(p'-1/2)$ and $z_q'=z_i-\Delta z/2+d_z'(q'-1/2)$, where $d_r'=\Delta r/n_r'$, $d_z'=\Delta z/n_z'$, and $(r_i,z_i)$ is the center of element $i$ cross-section. The number of elements in the $r$ and $z$ directions, $n_r'$ and $n_z'$, are determined in the same way as $n_r$ and $n_z$ for the ${\bf b}_j({\bf r})$ calculation above. Similarly, the number of sub-elements is increased until each component of ${\bf b}_{ij}$ satisfies a certain relative tolerance.

Although we may use the same process above for the computation of $C_{ij}$ in (\ref{cij}), in this work we use the numerical routine in \cite{pancaketheo}.


\section*{References}

\end{document}